\documentclass[twocolumn,nofootinbib,floatfix,superscriptaddress]{revtex4}

\usepackage{dcolumn}
\usepackage{graphicx}
\usepackage{rotating}
\usepackage{multirow}
\usepackage{amsmath,amsfonts,amssymb}

\newcommand{\g}{\,\mathrm{g}}
\newcommand{\kg}{\,\mathrm{kg}}

\newcommand{\remove}[1]{}

\begin{document}

\title{The evolution and distribution of species body size\footnote{This manuscript is a pre-print version that has not undergone final editing.  Please refer to the complete version of record, {\em Science} {\bf 321}, 399 -- 401 (2008), at {\tt http://www.sciencemag.org/}.  This manuscript may not be reproduced or used in any manner that does not fall within the fair use provisions of the Copyright Act without the prior, written permission of AAAS.}}
\author{Aaron Clauset}
\affiliation{Santa Fe Institute, 1399 Hyde Park Road, Santa Fe NM, 87501, USA}
\author{Douglas H. Erwin}
\affiliation{Santa Fe Institute, 1399 Hyde Park Road, Santa Fe NM, 87501, USA}
\affiliation{Departement of Paleobiology, MRC-121, National Museum of Natural History, P. O. Box 37012, Washington DC, 20013-7012, USA}

\begin{abstract}
The distribution of species body size within taxonomic groups exhibits a heavy right-tail extending over many orders of magnitude, where most species are significantly larger than the smallest species. We provide a simple model of cladogenetic diffusion over evolutionary time that omits explicit mechanisms for inter-specific competition and other microevolutionary processes yet fully explains the shape of this distribution. We estimate the model's parameters from fossil data and find that it robustly reproduces the distribution of 4002 mammal species from the late Quaternary. The observed fit suggests that the asymmetric distribution arises from a fundamental tradeoff between the short-term selective advantages (Cope's rule) and long-term selective risks of increased species body size, in the presence of a taxon-specific lower limit on body size.
\end{abstract}


\maketitle

\begin{figure}[b]
\begin{center}
\includegraphics[scale=0.4]{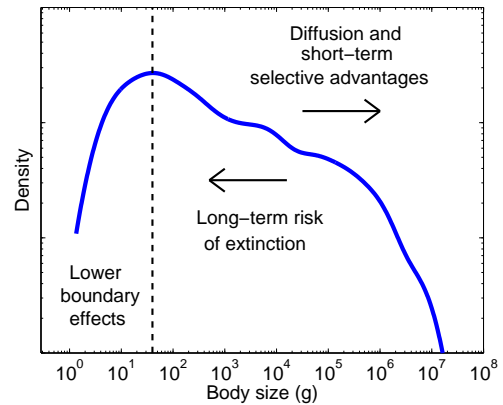} 
\end{center}
\caption{Smoothed species body size distribution of 4002 Recent terrestrial mammals (data from~\cite{smith:etal:2003}), showing the three macroevolutionary processes that shape the relative abundances of different sizes. The left-tail of the distribution is created by diffusion in the vicinity of a taxon-specific lower limit near $2\,\textrm{g}$, while the long right-tail is produced by the interaction of diffusion over evolutionary time (including trends like Cope's rule) and the long-term risk of extinction from increased body size.}
\label{fig:schematic}
\end{figure}

Most taxonomic groups show a common distribution of species body size~\cite{stanley:1973,kozlowski:gawelczyk:2002,allen:etal:2006}, with a single prominent mode relatively near but not at the smallest species size~\cite{dial:marzluff:1988} and a smooth but heavy right-tail (often described as a right-skew on a log-size scale) extending for several orders of magnitude (e.g., Fig.~\ref{fig:schematic}). This distribution is naturally related to a wide variety of other species characteristics with which body size correlates, including habitat, life history, life span~\cite{brown:1995}, metabolism~\cite{west:etal:2002} and extinction risk~\cite{cardillo:etal:2005}. A greater understanding of the underlying constraints on, and long-term trends in, body size evolution may provide information for conservation efforts~\cite{fisher:owens:2004} and insight about interactions between ecological and macroevolutionary processes~\cite{stanley:1975}.

Studies of body size distributions have suggested that the prominent mode may be indicative of a taxon-specific energetically optimal body size~\cite{sebens:1987,brown:etal:1996}, which is supported by microevolutionary studies of insular species~\cite{lomolino:1985}. However, evidence for Cope's rule~\cite{deperet:1909,stanley:1973,alroy:1998} -- the observation that species tend to be larger than their ancestors -- and the fact that most species are not close to their group's predicted optimal size (among other reasons~\cite{kozlowski:2002}), suggest that this theory may be flawed. Alternatively, species body sizes may diffuse over evolutionary time. If so, Cope's rule alone could cause size distributions to exhibit heavy right-tails~\cite{stanley:1973}, although size-dependent speciation or extinction rates~\cite{vanvalen:1973,stanley:1975,kozlowski:gawelczyk:2002} or size-neutral diffusion near a taxon-specific lower limit on body size~\cite{mcshea:1994} could also produce a similar shape. Furthermore, different mechanisms may drive body size evolution on spatial and temporal scales~\cite{allen:etal:2006}, and the importance of inter-specific competition to the macroevolutionary dynamics of species body size is not known.

\begin{figure}[t]
\begin{center}
\includegraphics[scale=0.55]{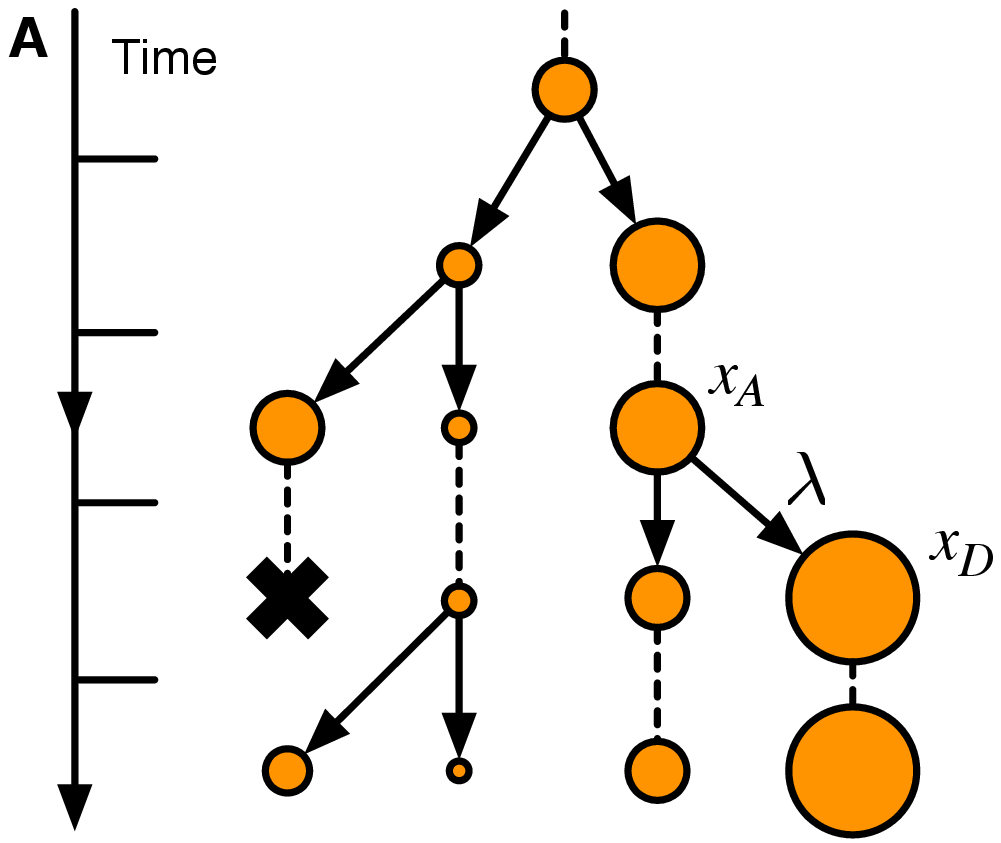} \\
\includegraphics[scale=0.35]{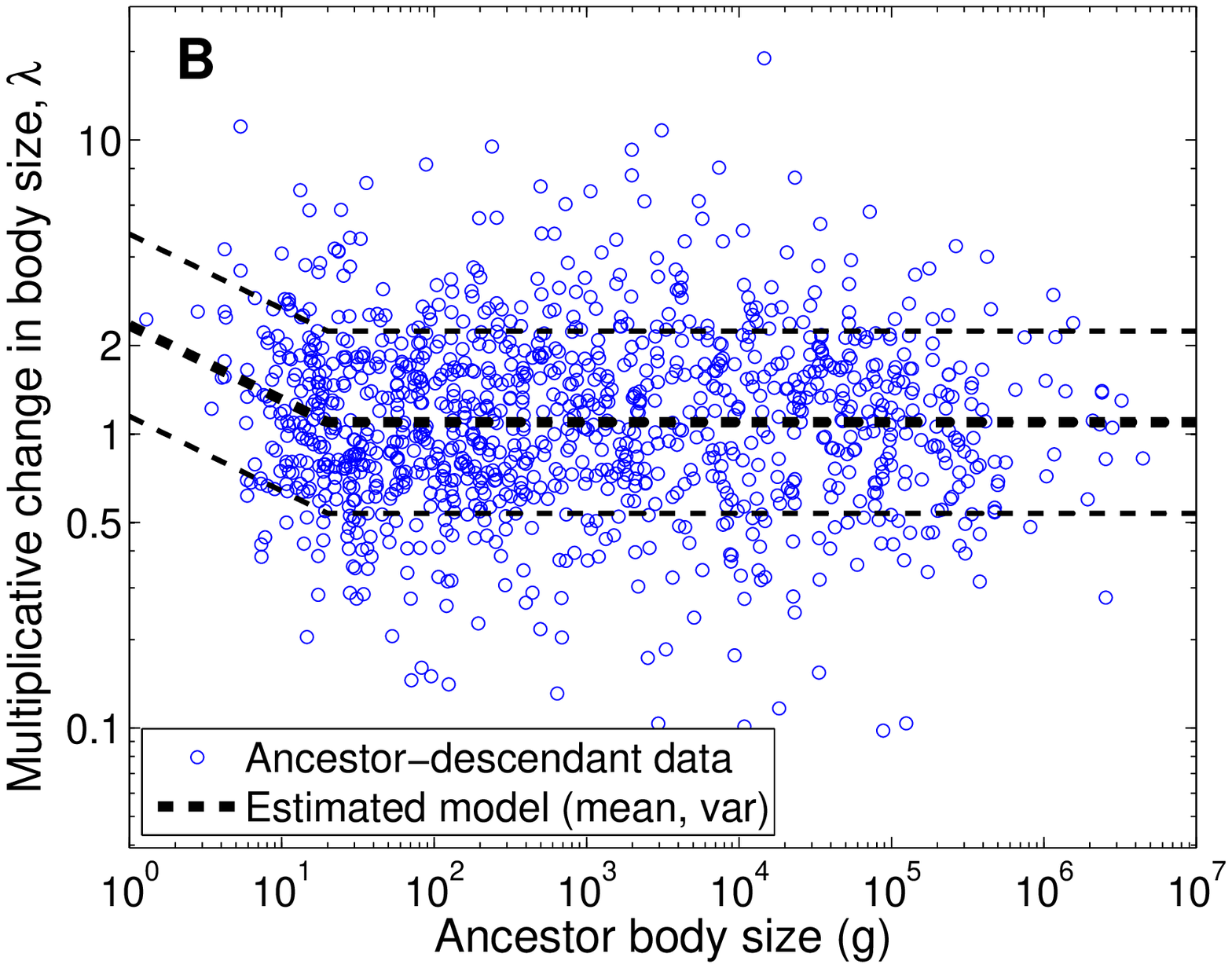}
\end{center}
\caption{(\textbf{A}) A schematic illustrating a simple cladogenetic diffusion model (see text) of species body size evolution, where the size of a descendant species $x_{D}$ is related to its ancestor's size $x_{A}$ by a multiplicative factor $\lambda$. (\textbf{B}) Empirical data on 1106 changes in North American mammalian body size (data from~\cite{alroy:2008}), as a function of ancestor size, overlaid with the estimated model of within-lineage changes, where the average log-change $\langle\log\lambda\rangle$ varies piecewise as a function of body size (see Appendix~\ref{appendix:copesrule:model}).} 
\label{fig:assumptions}
\end{figure}

We developed a generalized diffusion model of species body size evolution, in which the size distribution is the product of three macroevolutionary processes (Fig.~\ref{fig:schematic}). We combine these processes, each of which has been independently studied~\cite{stanley:1973,mckinney:1990,mcshea:1994,kozlowski:gawelczyk:2002}, in a single quantitative framework, estimate its parameters from fossil data on extinct terrestrial mammals from before the late Quaternary~\cite{fortelius:2003,alroy:2008}, and test whether this model, or simpler variants, can reproduce the sizes of the 4002 known extant and extinct terrestrial mammal species from the late Quaternary (Recent species)~\cite{smith:etal:2003}.

This model assumes that (1) species size varies over evolutionary time as a cladogenetic multiplicative diffusion process~\cite{stanley:1973,mcshea:1994}: the size of a descendant species $x_{D}$ is the product of a stochastic growth factor $\lambda$ and its ancestor's size $x_{A}$, i.e., \mbox{$x_{D}=\lambda\,x_{A}$}. For each speciation event, a new $\lambda$ is drawn from the distribution $F(\lambda)$, which models the total influence on species size changes from all directions. A bias toward larger sizes (Cope's rule) appears as a positive average log-change to size $\langle\log\lambda\rangle>0$, and may depend on the ancestor's size. \mbox{(2) Species} body size is restricted by a taxon-specific lower limit  $x_{\min}$~\cite{pearson:1948,west:etal:2002}, which we model by requiring that \mbox{$F(\lambda<x_{\min}/x_{A})=0$}, i.e., the largest possible decrease in size for a particular speciation event is \mbox{$\lambda=x_{\min}/x_{A}$}. In our computer simulations, time proceeds in discrete steps. At each step, exactly one new species is produced, which is the descendant of a randomly selected species. Finally, \mbox{(3) every} species independently becomes extinct with probability $p_{e}(x)$, which increases monotonically with size. A schematic of the model is shown in Fig.~\ref{fig:assumptions}A (for technical details see Appendix~\ref{appendix:model:specification}).

To make this model appropriately realistic, we estimated the form of each process from fossil data.
The lower limit on mammalian body size is near $2\,\textrm{g}$, close to the size of both the Etruscan shrew (\mbox{\emph{Suncus etruscus}}) and the bumblebee bat (\mbox{\emph{Craseonycteris thonglongyai}}). Fossil evidence suggests that this limit has existed since at least the Cretaceous-Tertiary boundary~\cite{fortelius:2003,smith:etal:2004,alroy:2008}. Further, a limit in this vicinity is supported by both experimental~\cite{pearson:1948} and theoretical work~\cite{west:etal:2002} on mammalian metabolism.

\begin{figure*}[t]
\begin{center}
\begin{tabular}{ccc}
\includegraphics[scale=0.317]{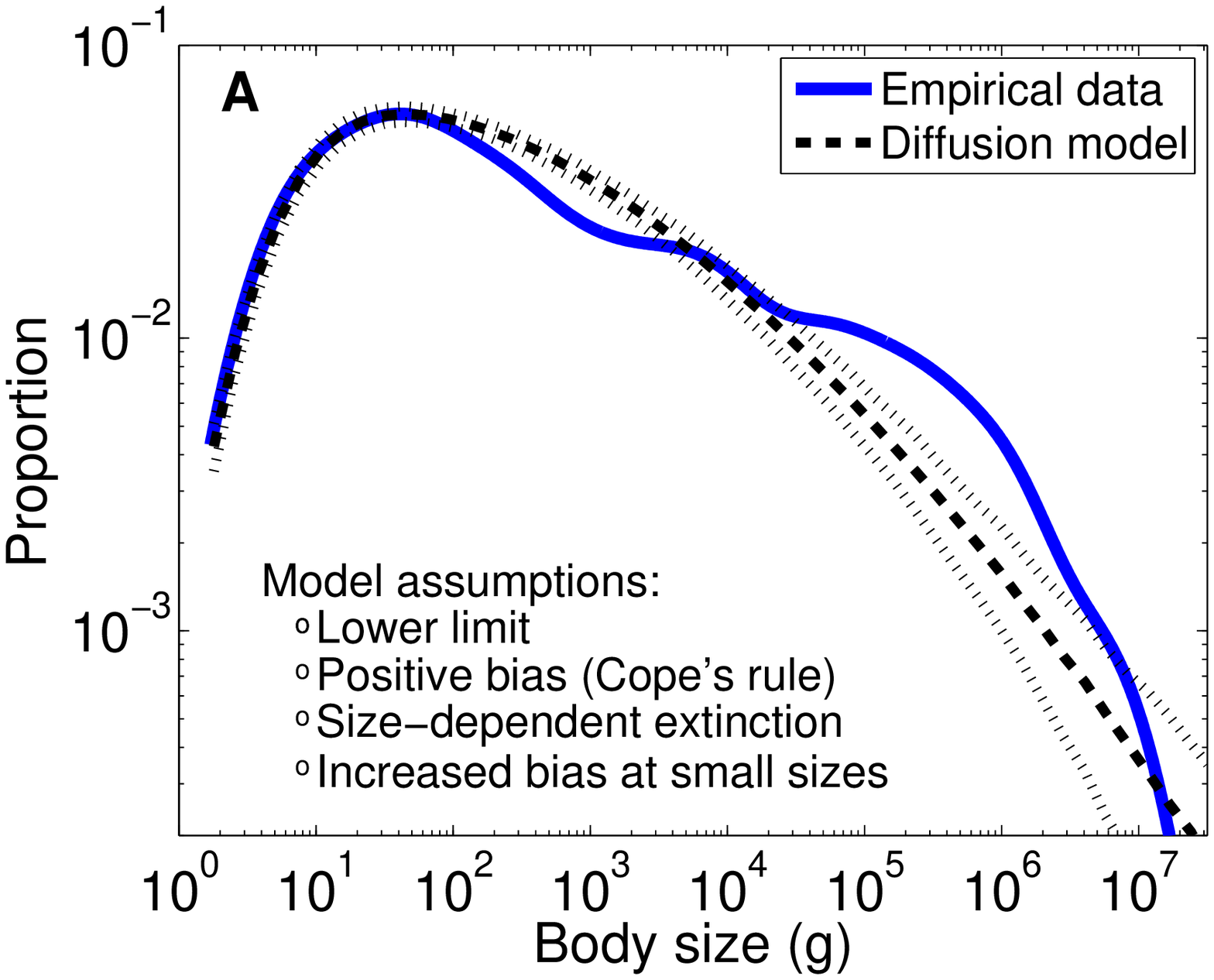} &
\includegraphics[scale=0.317]{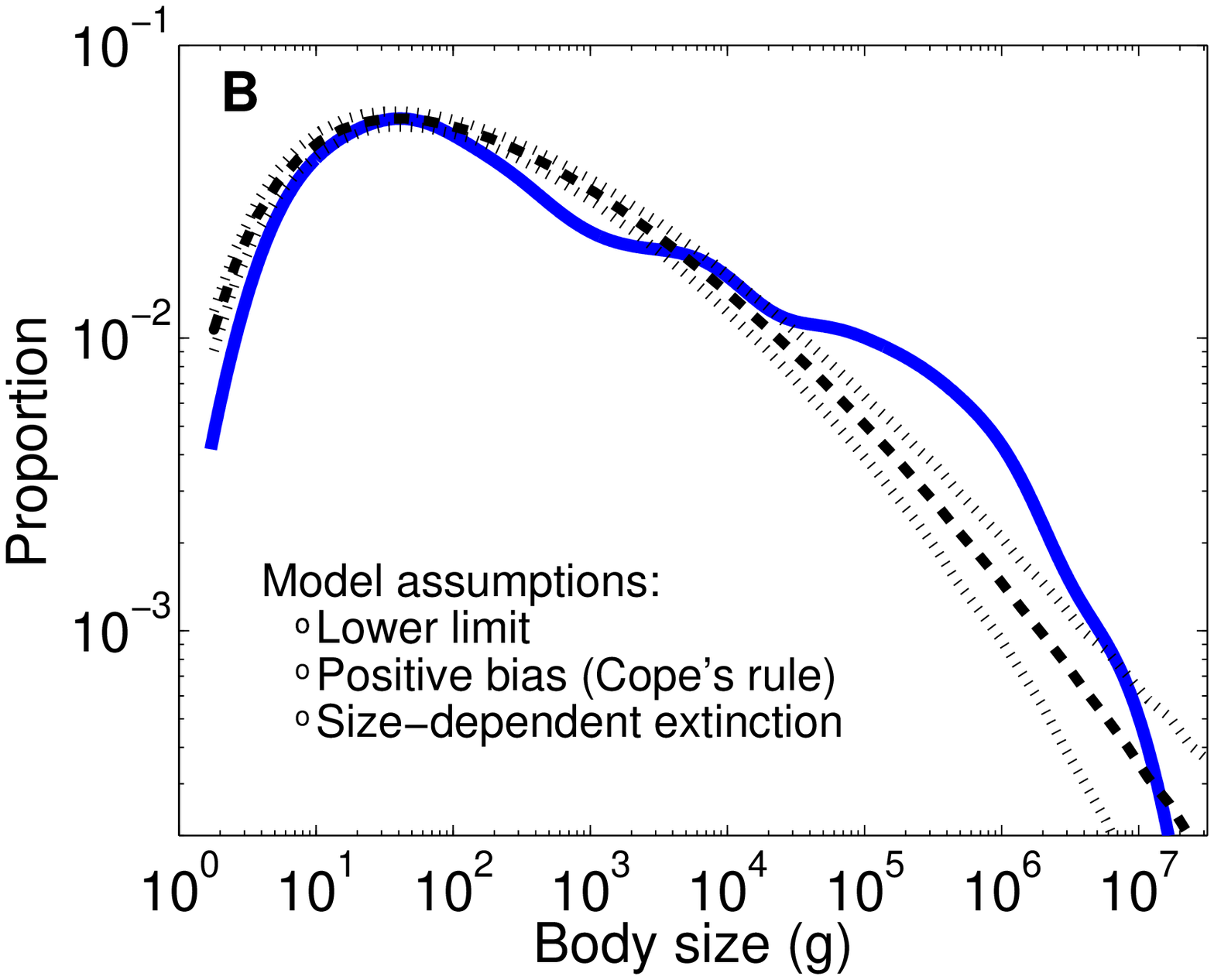} &
\includegraphics[scale=0.317]{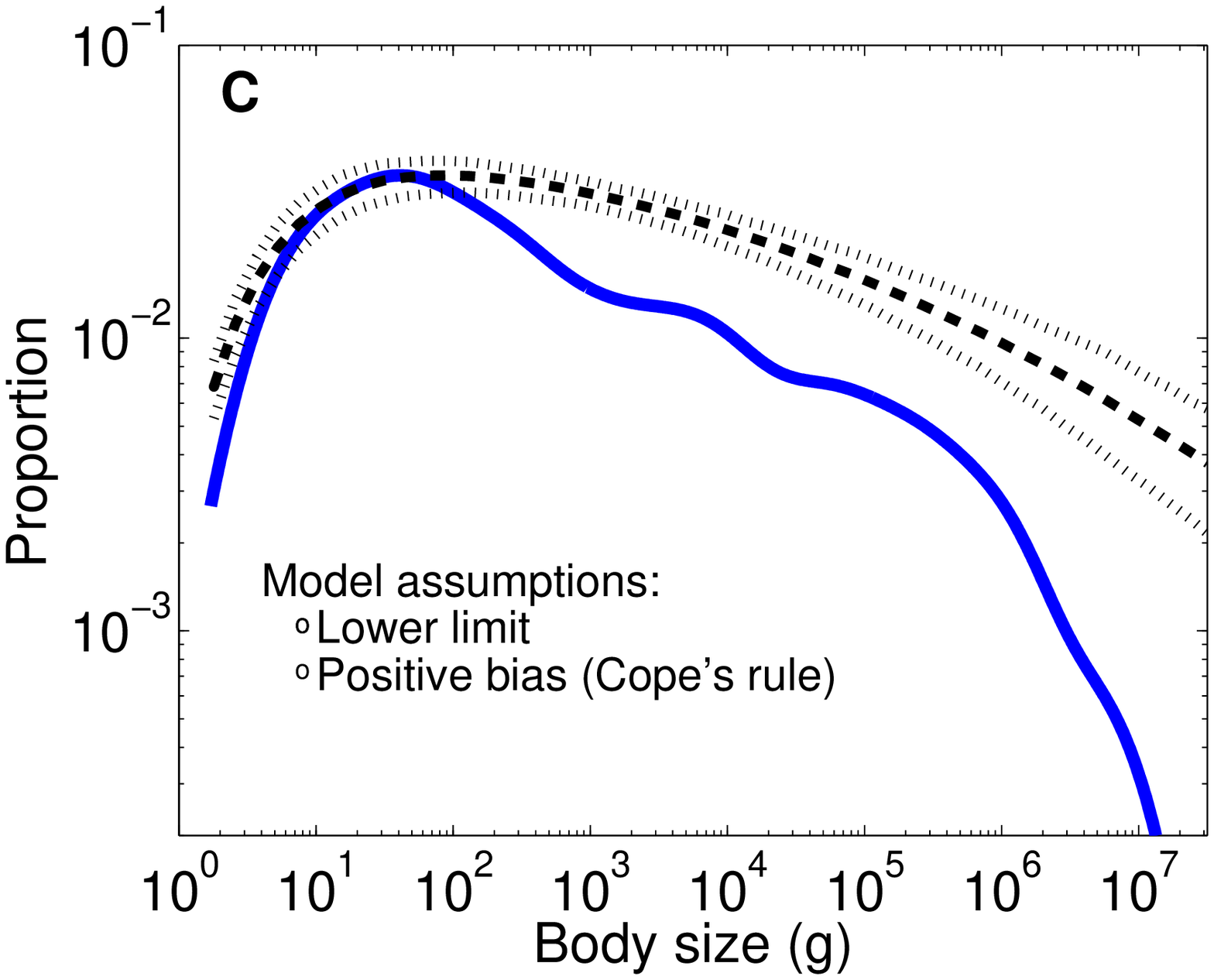} 
\end{tabular}
\end{center}
\caption{Simulated distributions of species body size (central tendency $\pm~95\%$ confidence intervals from 1000 repetitions; all model parameters estimated as described in the text) and the empirical distribution of Recent terrestrial mammals. (\textbf{A}) The model described in the text. (\textbf{B}) The same model as \textbf{A} but with a bias $\langle\log\lambda\rangle$ that is independent of size. (\textbf{C}) The same model as \textbf{B} but with an extinction risk that is independent of size. (For details and additional results, see Appendix~\ref{appendix:alternatives}.) } 
\label{fig:simulations}
\end{figure*}

Away from this limit, mammalian body size evolution is governed mainly by diffusion with a bias (Cope's rule)~\cite{alroy:1998,valkenburgh:etal:2004}, while its evolution near the lower limit is likely constrained by the need for relatively specialized morphological structures~\cite{stanley:1973}. We expect this latter effect to appear in fossil data as a systematic intensification of Cope's rule for very small-bodied species, i.e., increased $\langle\log\lambda\rangle$ as $x_{A}\rightarrow x_{\min}$. From ancestor-descendant size data for 1106 extinct North American terrestrial mammals~\cite{alroy:2008}, we estimated and compared three models of the distribution $F(\lambda)$ as a function of ancestor size, including the model suggested by Alroy~\cite{alroy:1998} which predicts a moderately bi-modal distribution in body sizes. Of these, a piecewise model (Fig.~\ref{fig:assumptions}B), with no effective optimal body size, has the best empirical support (model selection via likelihood ratio test and Bayesian information criterion; see Appendix~\ref{appendix:copesrule:model}). This model includes both a strengthening of Cope's rule for small-bodied species ($x\lesssim32\,\textrm{g}$) and a small but uniformly positive bias for larger species, resulting in an average body-size growth of $4.1\pm1.0\%$ between ancestors and their descendants ($\langle\log\lambda\rangle=0.04\pm0.01$).

This result supports the existence of short-term selective advantages for increased species body size, e.g., better tolerance of resource fluctuations, better thermoregulation, and better predator avoidance~\cite{brown:1995}, but also implies a more nuanced view: small-bodied species exhibit even greater selective advantages from increased size, e.g., because of greater morphological flexibility. 

Empirical estimates of extinction rates (or equivalently, speciation rates) as functions of body size are uncertain~\cite{ludwig:1996}, due to the bias and incompleteness of the fossil record. We partly control for this uncertainty by utilizing a simplistic model of extinction risk $p_{e}(x)$, largely estimated from the data, where extinction occurs independently with a probability calculated only from the species' size. We specified a basal extinction rate $\beta$ by assuming that the number of Recent terrestrial mammal species is close to a putative carrying capacity. We then let extinction risk per unit time increase logarithmically with body size~\cite{liow:etal:2008} (see Appendix~\ref{appendix:model:estimation}). This model leaves only the rate $\rho$ by which risk increases with size as a free parameter, which was chosen by minimizing the statistical distance between the simulated and empirical distributions (see Appendix~\ref{appendix:model:scoring}).

Inserting these three processes, as estimated above, into our computer model, we found that the model accurately predicted the distribution of Recent terrestrial mammal sizes over its seven orders of magnitude (Fig.~\ref{fig:simulations}A), and was particularly accurate for small-bodied species ($x<80\,\textrm{g}$). 
Our sensitivity analysis further indicated that this prediction was highly robust to variations in most of the estimated parameters, but highly sensitive to the location of the lower-limit on body size. The estimated value of $x_{\min}\approx2\,\textrm{g}$, however, is the most strongly supported of all model parameters. Thus, even large revisions to the other parameter estimates are unlikely to change our general conclusions (see Appendix~\ref{appendix:sensitivity}).
Also, although a range of $\rho$ values produced size distributions that were statistically close to the empirical distribution, the model predicts a particular extinction risk curve (Fig.~~\ref{fig:sensitivity1}) that could be tested with appropriate empirical data.

To further discriminate among alternative explanations for the species size distribution, we tested simpler diffusion models, each with parameters estimated from fossil data (see Appendix~\ref{appendix:alternatives}), including (1) unbiased diffusion with a lower boundary, (2) Cope's rule with size-dependent extinction, (3) Cope's rule alone, (4) size-dependent extinction alone, and (5) a version of the full model that omits the increased bias for small-bodied species ($x\lesssim32\,\textrm{g}$). We found that these models all predicted size distributions that differed, sometimes dramatically so, from the empirical distribution (Figs.~\ref{fig:simulations}B,~\ref{fig:simulations}C,~\ref{fig:comparison:1} and~\ref{fig:comparison:2}). Additionally, we found that a positive bias $\langle\log\lambda\rangle>0$ for large-bodied species is not necessary if the extinction risk increases less quickly (see Appendix~\ref{appendix:sensitivity}). These results support the inclusion of a fundamental lower limit, the diffusion of species size, and an increasing risk of extinction with size, as well as an increased bias toward larger sizes for small-bodied species ($x\lesssim32\,\textrm{g}$).

Thus, the shape of a body size distribution can be interpreted in the context of these three macroevolutionary processes. An intermediate location for the distribution's mode ($40\,\textrm{g}$ for terrestrial mammals) is mainly caused by diffusion in the vicinity of the physiological lower limit on body size -- which prevents the smallest species from being the most abundant.  A heavy right-tail is then caused primarily by diffusion in the presence of extinction risks that increase weakly with size $\rho>0$. For mammals, the within-lineage tendency toward increased size (Cope's \mbox{rule, $\langle\log\lambda\rangle>0$)} shifts the mode toward slightly larger sizes and slightly increases the heaviness of the right-tail.

Under different conditions, these processes produce markedly different body size distributions. For instance, a long left-tail extending toward small-bodied species would indicate that the risk of extinction decreases with larger size $\rho<0$. Similarly, a more symmetric distribution would indicate both that extinction rates are relatively size-independent $\rho\approx0$ and that changes to body size convey few selective advantages $\langle\log\lambda\rangle\approx0$. Although a suitable body size distribution is not currently available for dinosaurs (but see~\cite{carrano:2006}), evidence suggests that it may be more symmetric than for mammals. The right-skewed distribution's ubiquity, such as for insects and birds~\cite{stanley:1973,kozlowski:gawelczyk:2002}, suggests that such circumstances are rare, and that the mammalian distribution represents the norm.

This model omits explicit mechanisms for many canonical ecological and microevolutionary processes, including the impact of inter-specific competition, geography, predation, population dynamics, and size variation between speciation events (anagenetic evolution), which suggests that their contributions to the systematic or large-scale character of species body size distributions can be compactly summarized by the values of certain model parameters, e.g., the strength of Cope's \mbox{rule $\langle\log\lambda\rangle$} or the manner in which extinction risk increases with body \mbox{size $\rho$}. Some aspects of the body size distribution, however, are not explained by this model, such as the slight over-abundance of terrestrial mammal species around $300\,\textrm{kg}$ and the slight under-abundance around $1\,\textrm{kg}$ (Fig.~\ref{fig:simulations}A). Whether such deviations can be attributed to phylogenetically correlated speciation and extinction events is an open question.
A more thorough examination of these macroevolutionary processes may explain their particular form and origin, and answer why body size is weakly correlated with increased extinction rates (or, decrease of speciation rates) weakly with body size, why physiological lower limits on body size exist and are conserved within a taxonomic groups, and why some groups exhibit macroevolutionary trends but others do not.

\begin{acknowledgments}
AC is grateful to A. Boyer, J. Dunne, J. Ladau, B. Olding, C. Shalizi, and J. Wilkins for helpful conversations. We thank J. Alroy, A. Boyer and F. Smith for kindly sharing data. Supported in part by the Santa Fe Institute and the Computer Science Department at the University of New Mexico.
\end{acknowledgments}



\renewcommand{\thefigure}{S\arabic{figure}}
\setcounter{figure}{0}
\renewcommand{\thetable}{S\arabic{table}}
\setcounter{table}{0}

\newpage

\begin{appendix}
\section*{Appendices}
\label{appendix:start}
These appendices document the technical details of our study.
\begin{itemize}
\item Appendix~\ref{appendix:model} fully describes the cladogenetic model used to test our main hypotheses, including the model's specifications (Appendix~\ref{appendix:model:specification}), the statistical estimation of the model parameters from the mammalian fossil data (Appendix~\ref{appendix:model:estimation}), and our score function for comparing the results of the model to empirical data (Appendix~\ref{appendix:model:scoring}).
\item Appendix~\ref{appendix:copesrule} describes our model of species size variation at speciation events, including a new analysis of the empirical evidence for Cope's rule (Appendix~\ref{appendix:copesrule:evidence}) and the estimation of the distribution $F(\lambda)$ of within-lineage changes to body size (Appendix~\ref{appendix:copesrule:model}).
\item Appendix~\ref{appendix:model:additional} presents supplementary results from simulating the model, including snapshots from a single simulation (Appendix~\ref{appendix:model:results}), and the results of our analysis of the model's sensitivity to the estimated parameters (Appendix~\ref{appendix:sensitivity}).
\item Appendix~\ref{appendix:alternatives} presents detailed comparisons of the model with simpler alternative diffusion models, several of which have previously been suggested as explanations of right-skewed size distributions.
\item Appendix~\ref{appendix:code} gives a complete Matlab-code implementation of the model.
\end{itemize}

\section{A cladogenetic diffusion model of body size evolution}
\label{appendix:model}
Complex theoretical questions about the evolution of body size, such as the ones we consider, are typically explored with simulations. Such a choice is mainly driven by the fact that a mathematical analysis of branching processes is often intractable for all but the most simple questions. On the other hand, poorly executed simulation studies can be misleading as a result of incorrect specification, among other reasons. We make a concerted effort to avoid such problems by defining a model whose parameters can be estimated directly from fossil data prior to the late Quaternary, and whose output can be validated against data from the late Quaternary (Recent species). Although these two data sources are not logically independent, they are perhaps as close to independent as we might wish for such a macroevolutionary study. We note that while we mainly study the body size distribution of terrestrial mammals here, this framework can easily be adapted to other taxonomic groups, e.g., birds.

\subsection{Model specification}
\label{appendix:model:specification}
As described in the main text, our model combines three simple mechanisms related to body size evolution. Each of these processes has been previously suggested or studied the literature, but are combined here in a coherent, quantitative framework that engages directly with empirical data.  We now briefly describe the technical details of the three processes.
\begin{enumerate}
\item The range of possible body sizes for a particular higher taxon, e.g., terrestrial mammals, obeys a lower limit $x_{\min}$. A limit like this was suggested in~\cite{stanley:1973} on the basis that physiological factors, e.g., metabolic requirements, constrain how small a particular body plan can become without fundamental innovation. (For convenience, we also assume that body size obeys an upper limit, but set this limit at an extremely large size, $x_{\max}=10^{15}\g$.)

\item As is conventional, simulated time proceeds in discrete steps, each of which corresponds to a single event of cladogenesis. Although realistically, each cladogenetic event could produce a variable number of descendent species, we present results only for the case where exactly two new species are created while the ancestor species becomes extinct. We note that several apparently reasonable variations on this rule, e.g., creating one or more descendent species while letting the ancestral species continue, however, appear to produce equivalent results.

At each of these speciation events, each descendent species' body size $x_{D}$ varies from its ancestor's body size $x_{A}$ according to a multiplicative random walk. That is, the size of a descendent is the product of its ancestor's body size and a random variable $\lambda$, which represents the relative percentage change in body size due to all contributing factors. We then assume that the instantaneous distribution of changes to body size $F(\lambda)$ for a given event has two main characteristics: (1) it is stable over evolutionary time (i.e., it is not a function of time $t$, although it may be a function of ancestor size $x_{A}$), and (2) it always respects the aforementioned limits on body size.  This latter requirement implies that for a given ancestor body size $x_{A}$, the distribution of allowed changes to size $F(\lambda)$ is bounded on the interval $[\frac{x_{\min}}{x_{A}}, \,\frac{x_{\max}}{x_{A}}]$. Fig.~\ref{fig:appendix:model}B illustrates this idea, showing how the support of the distribution varies as a function of body size. If $\langle\log\lambda\rangle\not=0$, then we say that $F(\lambda)$ is ``biased,'' with a positive bias corresponding to Cope's rule; if $\langle\log\lambda\rangle=0$, we say that $F(\lambda)$ is ``unbiased.''

\begin{figure*}[t]
\begin{center}
\begin{tabular}{cc}
\includegraphics[scale=0.531]{model_schematic_A.eps} &
\includegraphics[scale=0.35]{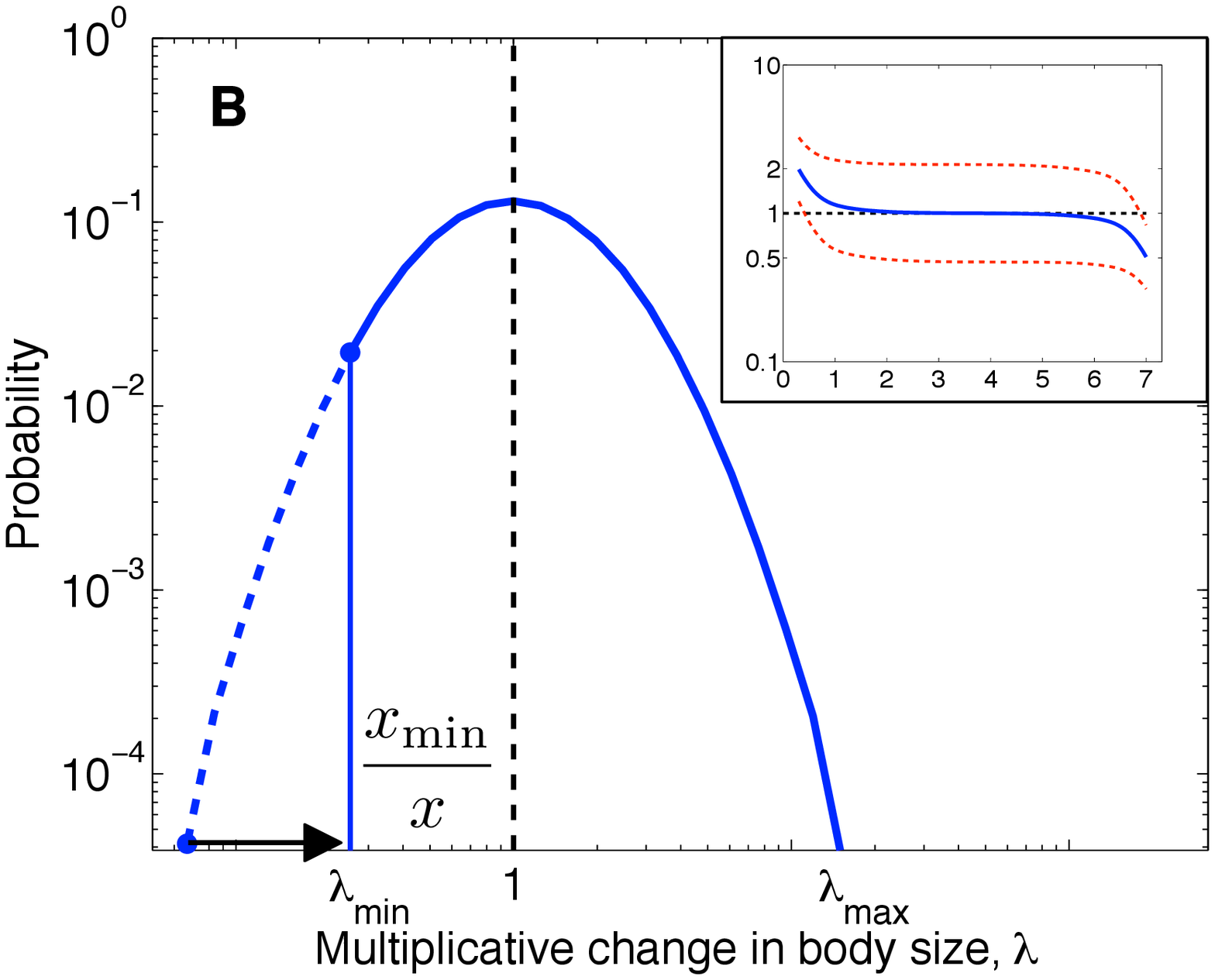} \\
\includegraphics[scale=0.35]{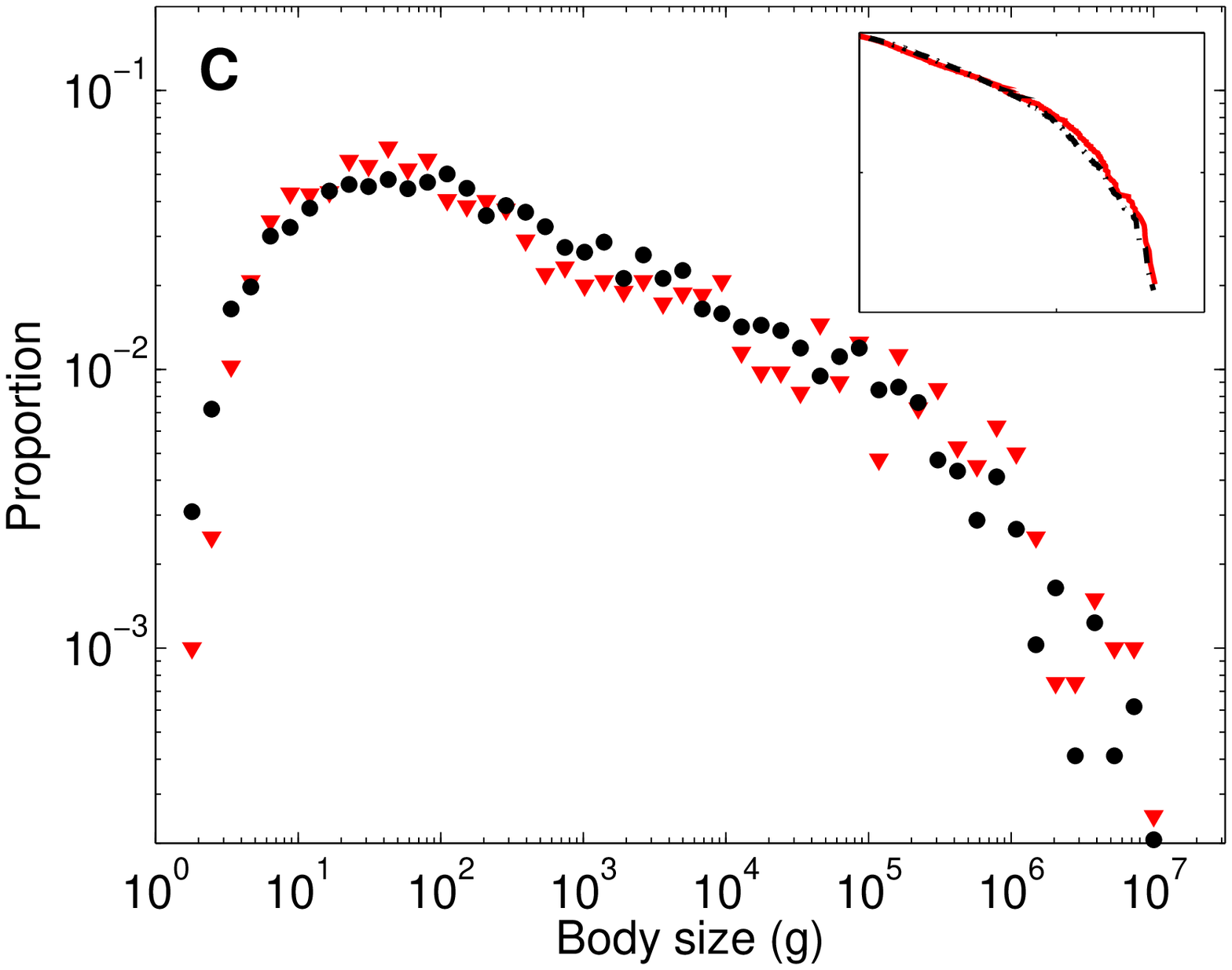} &
\includegraphics[scale=0.35]{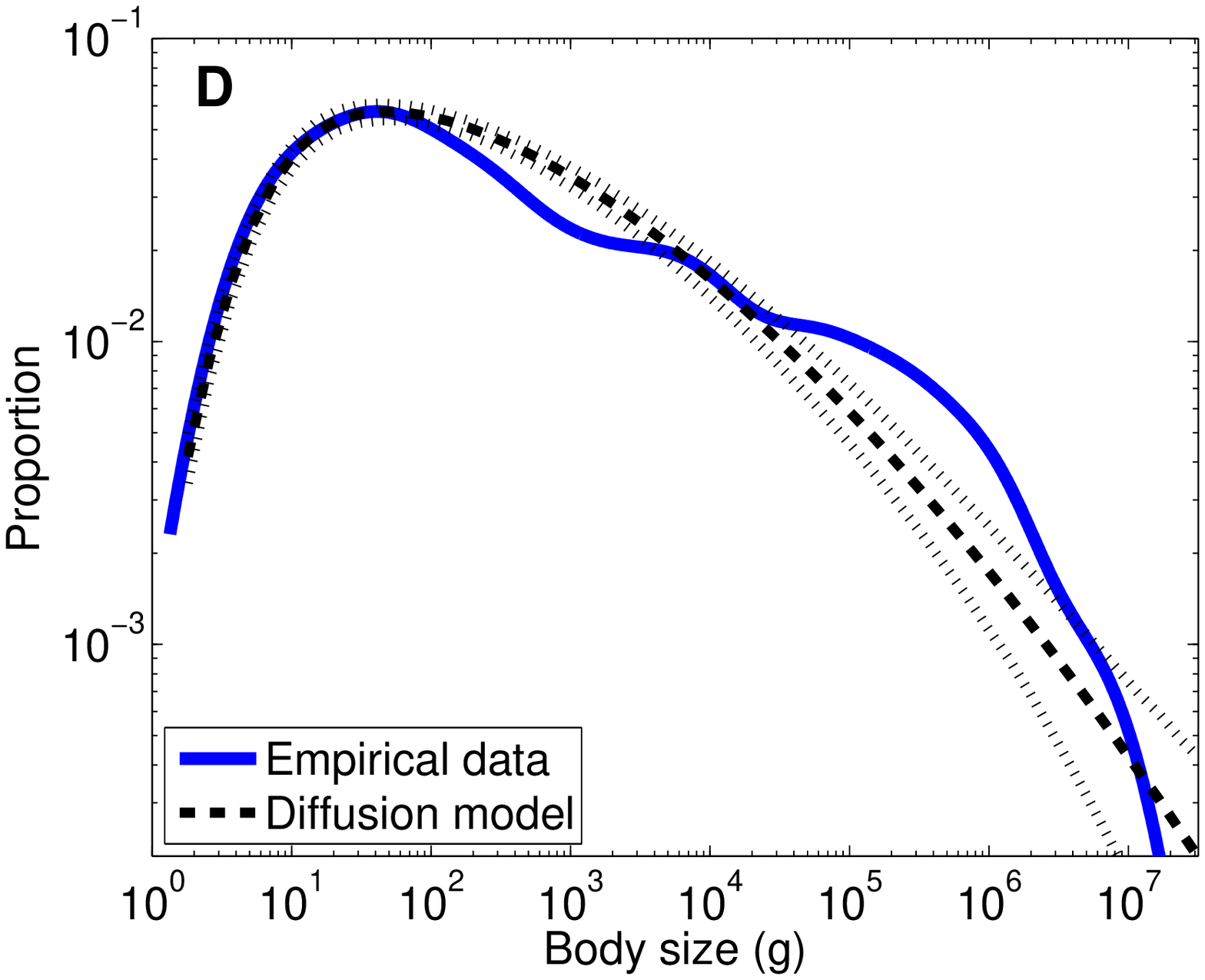} 
\end{tabular}
\end{center}
\caption{Results of modeling the evolution of body sizes for 4002 Recent terrestrial mammal species. (\textbf{A}) A schematic illustrating a simple cladogenetic model (see text) of species body size evolution, on the basis of a multiplicative diffusion process where the size of a descendant species $x_{D}$ is related to its ancestor's size $x_{A}$ by a multiplicative factor $\lambda$. (\textbf{B}) Model of the distribution of within-lineage body size changes $F(\lambda)$, where lower and upper boundaries on body size are enforced by letting setting \mbox{$F(\lambda<x_{\min}/x) = 0$}. Thus, as a lineage approaches $x_{\min}$, the distribution increasingly favors changes in the opposite direction of the limit (inset: average change in log-body size, as a function of ancestral body size, with $\langle\log\lambda\rangle=0$, $x_{\min}=1.8\g$ and $x_{\max}=10^{7}\g$). We incorporate a model of Cope's rule by letting the mean of this distribution $\mu(x_{A})$ vary as a function of $x_{A}$, where $\mu(x_{A})$ is estimated from fossil data (see Appendix~\ref{appendix:copesrule}). (\textbf{C}) Histogram of Recent mammal body sizes overlaid by an example distribution produced by the model (inset: corresponding complementary cumulative distribution functions). (\textbf{D}) The central tendency (with 95\% confidence intervals) of the simulated distribution of species body sizes and the smoothed empirical distribution for 4002 Recent mammal species (Gaussian kernel).} 
\label{fig:appendix:model}
\end{figure*}

In the physics literature (see~\cite{boas:2006}), this boundary effect is similar to an ``absorbing boundary'' condition in a diffusion-reaction equation, i.e., we require that the probability density go to zero at the boundary, $s(x)=0$ at $x=x_{\min}$. In contrast, a ``reflecting'' or ``insulating boundary'' would require that the flux across the boundary be zero, ${\rm d}s/{\rm d}x=0$ at $x=x_{\min}$. Unfortunately, these same terms have different meanings in the body size literature (see~\cite{mcshea:1994}); thus, we avoid their use entirely.

\item Species become extinct independently with a probability $p_{e}$ that depends only on species body size. We considered two functional forms for how this risk of extinction varies with body size: a power-law function of the form \mbox{$\log_{10}\,p_{e}(x)=\rho \log_{10} \,x + \log_{10} \,\beta$}, where $\beta$ is the baseline extinction rate and $\rho$ is the rate at which the rate increases with log-body size, and a logarithmic function \mbox{$p_{e}(x)=\rho \log_{10} \,x + \beta$}.

\begin{figure*}[t]
\begin{center}
\begin{tabular}{ccc}
\includegraphics[scale=0.315]{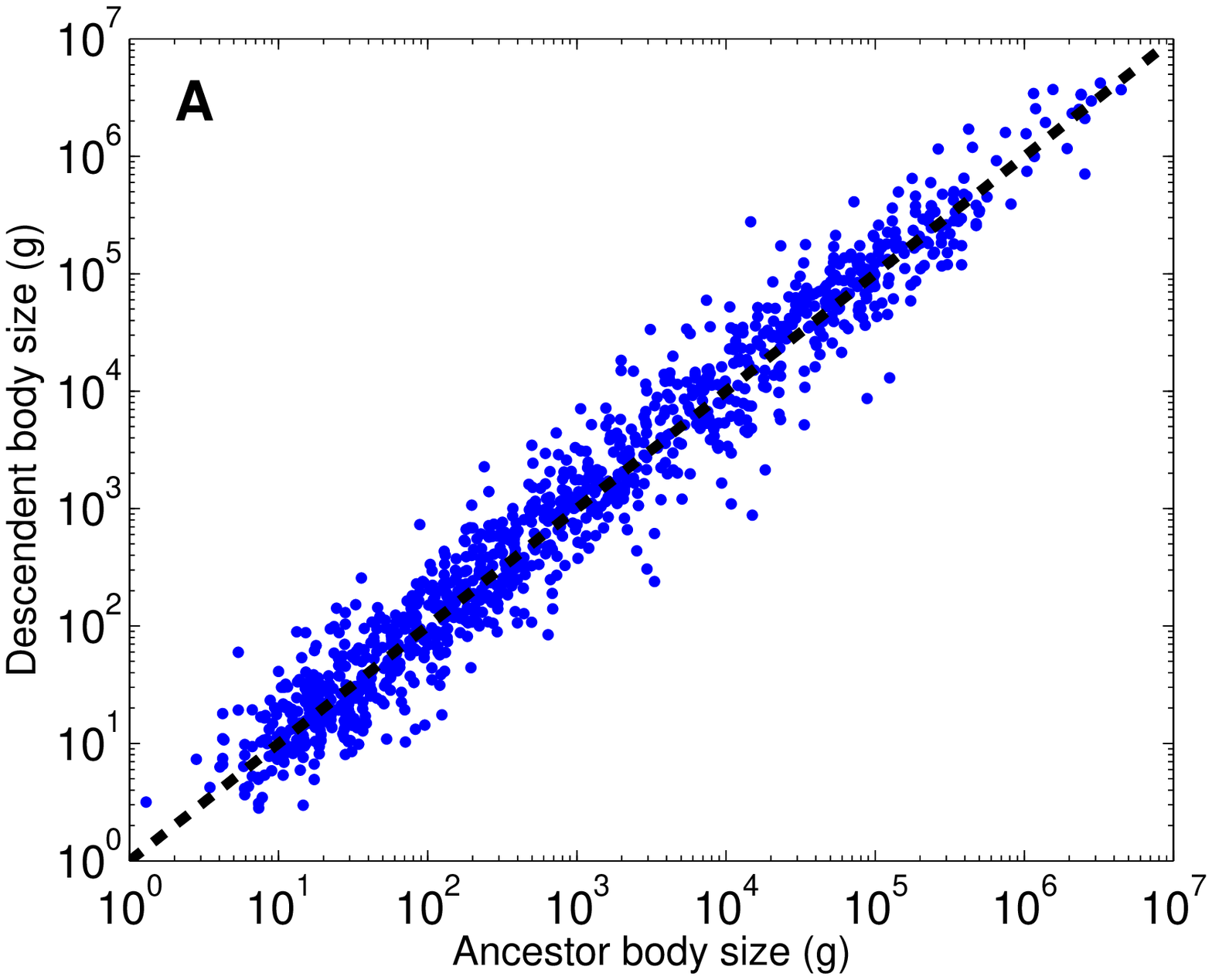} &
\includegraphics[scale=0.315]{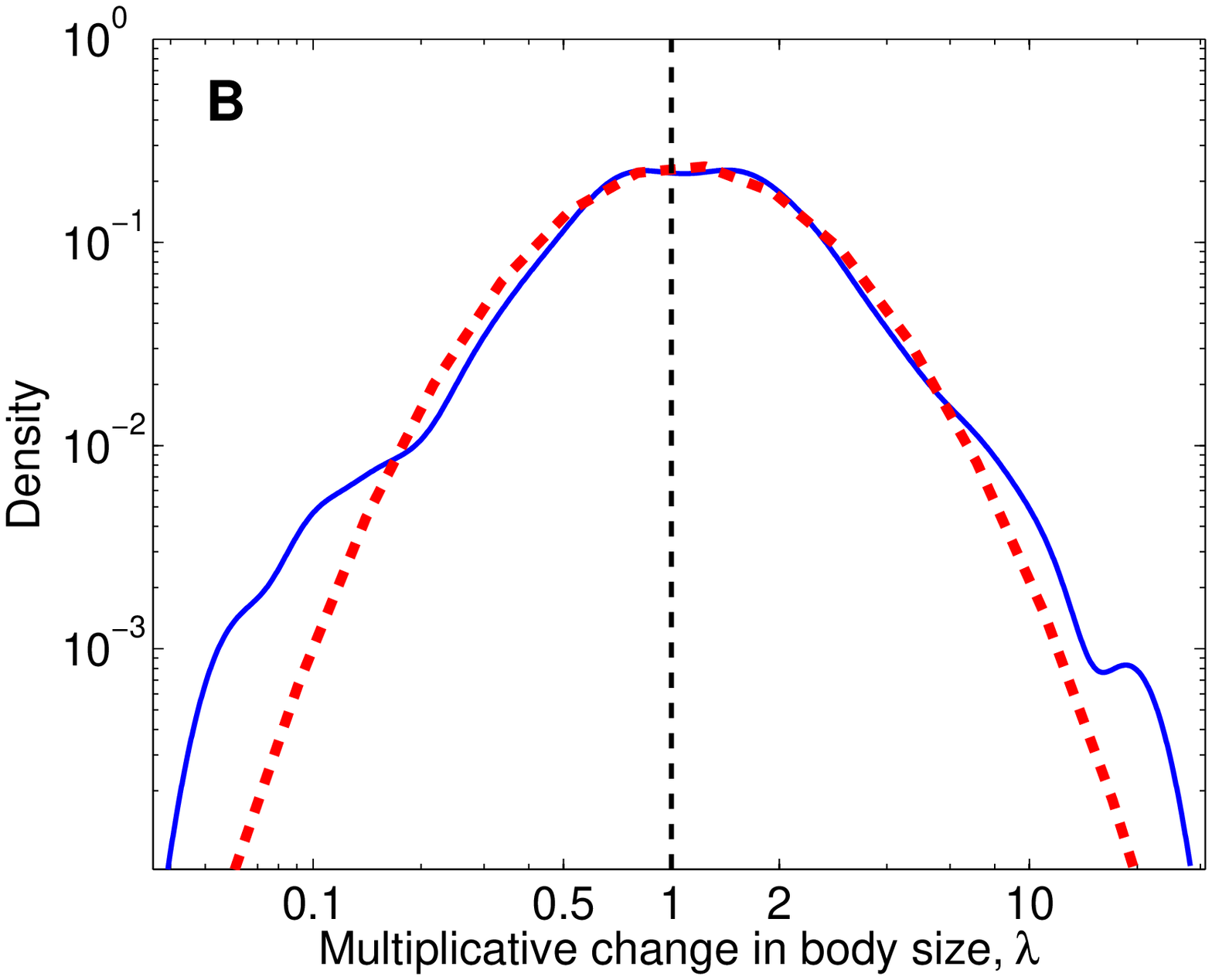} &
\includegraphics[scale=0.315]{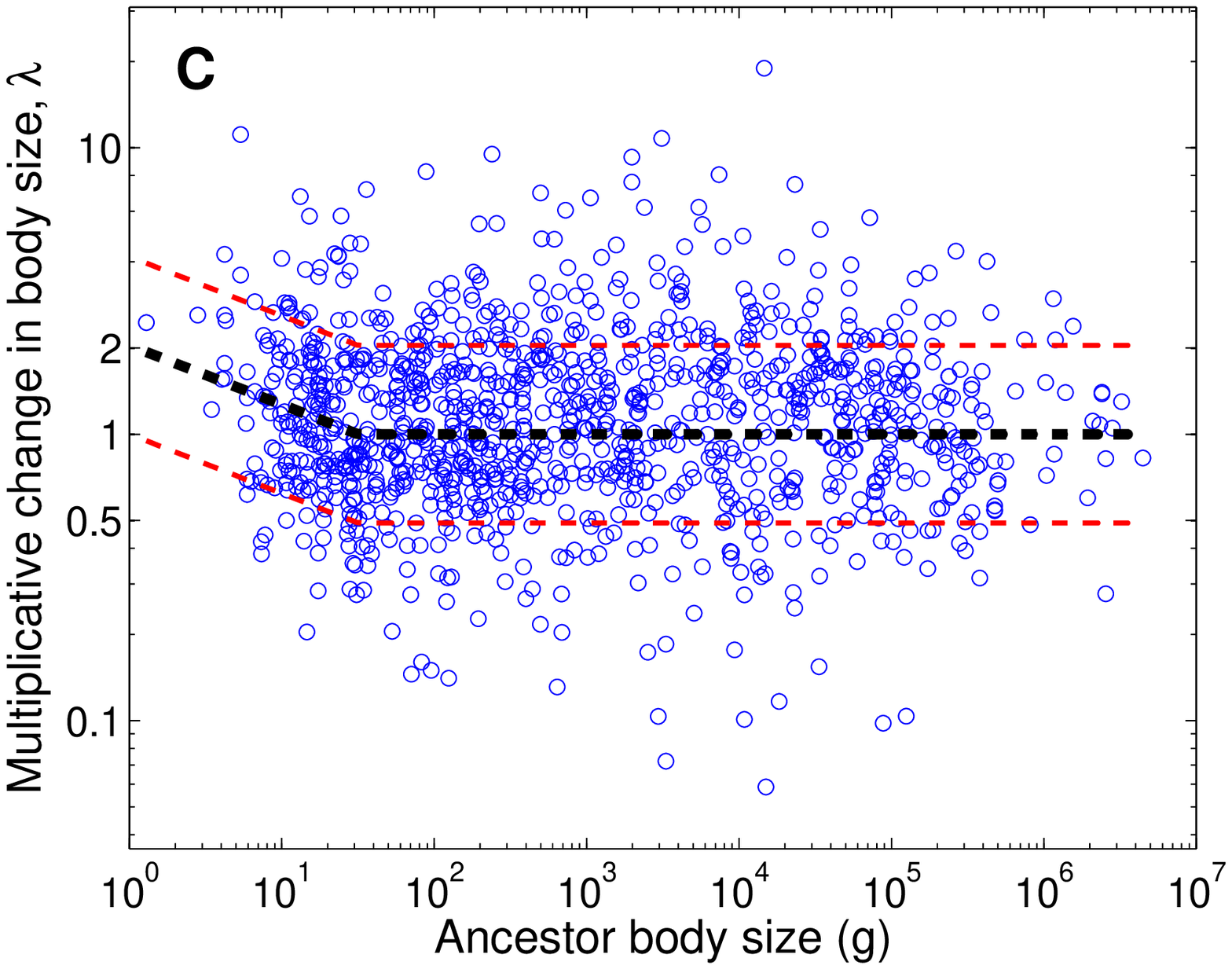} 
\end{tabular}
\end{center}
\caption{Analysis of 1106 pairs of mammal species in the North American fossil record~\cite{alroy:2008}. (\textbf{A}) Descendent body size $x_{D}$ versus ancestor body size $x_{A}$ overlaid by the relation \mbox{$x_{D}=x_{A}$}, representing the null-hypothesis of no bias toward larger or smaller body sizes, i.e., $\langle\log\lambda\rangle=0$. The best-fit allometric relation $\log x_{D}=\tilde{\lambda}\log x_{A}$ for this body size data (by standardized major axis regression~\cite{warton:etal:2006}) produces an estimated slope $\tilde{\lambda}=1.02\pm0.1$ (where $\pm$ indicates the 95\% confidence interval; $r^{2}=0.95$). (\textbf{B}) Estimated density (Gaussian kernel) of the distribution $F(\lambda)$ of within-lineage changes to species body size (solid line; equivalent to distribution of vertical residuals \mbox{in \textbf{A}),} along with the maximum likelihood log-normal distribution (dashed). (\textbf{C}) Change in species body size $\lambda$ as a function of ancestor size (circles) overlaid with the best model of the form $\log \lambda(x_{A})=\mathcal{N}[\mu(x_{A}),\sigma^{2}]$ (dashed lines). Under this model, changes in body size at speciation events are systematically biased toward larger sizes (Cope's rule); the bias is strongest for small bodied species, but still positive [$\mu(x_{A})=0.04$] for larger species $x\gtrsim32\g$. A likelihood ratio test indicates that this model is a better fit to the data than a model with no bias [$\mu(x_{A})=0$] for larger species ($p=1.44\times 10^{-4}$; see Appendix~\ref{appendix:copesrule:model}). We note that this simple model is a more conservative one than a model that includes the heavy tails of the distribution shown in \textbf{B}.}
\label{fig:paleo}
\end{figure*}

The notion that extinction risk increases with body size $\rho>0$ is a conventional one in the body size literature~\cite{liow:etal:2008}, although most empirical documentation of these notions concern relatively modern species. As such, relatively little is known about speciation and extinction rates in the fossil record~\cite{ludwig:1996,erwin:2006}. However, as population size generally decreases with increased body size, the increased extinction risk could result from populations of larger sized organisms being closer to inviable population sizes. The result for this mechanism is that one parameter -- the rate at which extinction risk increases with body size $\rho$ -- remains free in our study.

We note that an equivalent model would allow the speciation rate, or both extinction and speciation, to vary with body size. The absolute value of the speciation and extinction rates is not important~\cite{kozlowski:gawelczyk:2002}, but rather their ratio is. For a discrete-time model, size-dependent extinction rates are significantly easier to work with.
\end{enumerate}
Only a few more words are necessary to complete our specification of the model. At each time step, one species, chosen uniformly at random from the extant set, undergoes cladogensis according to Rule 2. This action produces two daughter species, one of which is new and the other of which replaces the ancestral species in the extant set. Subsequently, each extant species becomes extinct according to Rule 3; extinct species are removed from the extant set. Fig.~\ref{fig:appendix:model}A illustrates this branching process schematically. The model is initialized with a single founder species with body size $x_{0}$, and proceeds for $t_{\max}$ time steps (the number of steps is also the cumulative number of species produced). Fig.~\ref{fig:appendix:model}B illustrates the form of Rule 2 that we use (see Appendix~\ref{appendix:copesrule} for more details), where the largest change in body size is constrained so that the result would be to produce a daughter species with size $x_{\min}$. Fig.~\ref{fig:appendix:model}C shows an example of the resulting simulated distribution of species body sizes, where we have used the parameter values given in Table~\ref{table:params}, and Fig.~\ref{fig:appendix:model}D shows the central tendency of this model.

\subsection{Parameter estimation}
\label{appendix:model:estimation}
To implement this model on a computer, we must choose the form of each mechanism, e.g., $F(\lambda)$. Where possible, we estimated both the form and the corresponding parameters directly from fossil data; the only genuine free parameter in the model is $\rho$, the rate at which extinction risk increases with size. In this section, we describe our methodology for estimating parameters for Rules 1 and 3, the size of the founder species, and the number of species to simulate. The methodology for parameterizing Rule 2 is slightly more involved and is described subsequently (Appendix~\ref{appendix:copesrule}).

Rule 1 (boundaries) requires parameters to define a lower limit on body size. The most direct way to estimate these values is to consider fossil~\cite{alroy:2008,fortelius:2003} and Recent~\cite{smith:etal:2003} body size data. Each of these sources agrees that the minimum mammalian body size is in the neighborhood of $x_{\min}\approx2\g$ [e.g., both the Etruscan shrew (\emph{S. etruscus}) and the bumblebee bat (\emph{C. thonglongyai}) are in this range]. Experimental~\cite{pearson:1948} and theoretical work~\cite{west:etal:2002} on metabolism also supports a fundamental limit in this vicinity. The particular size of the founder species has little impact on the simulation results (see Appendix~\ref{appendix:sensitivity}), and for convenience we choose it to be equal to the mode of the Recent distribution, $x_{0}=40\g$. 

\begin{figure}[t]
\begin{center}
\includegraphics[scale=0.45]{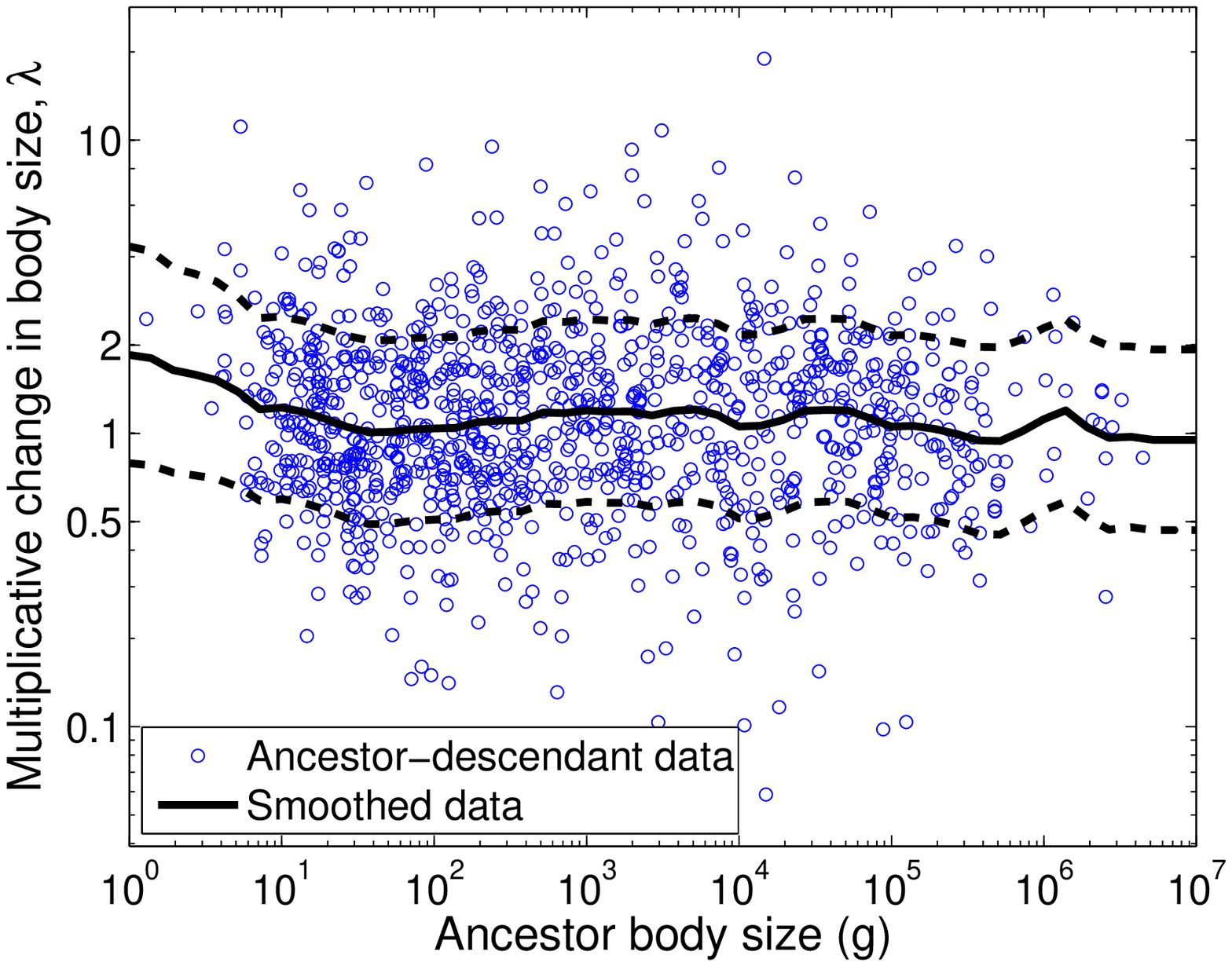} 
\end{center}
\caption{The same data as in Fig.~\ref{fig:paleo}C along with a smoothed version (exponential kernel) showing the mean $\pm$ one standard deviation. The smoothed trend is quite similar to the piece-wise linear model that we fitted to the data via maximum likelihood (see Appendix~\ref{appendix:copesrule:model}).}
\label{fig:smoothed}
\end{figure}

Parameter estimates for Rule 3 (extinction rates) can be partially derived from existing fossil data. We estimate the baseline extinction rate $\beta$ for terrestrial mammals in the following way. If the number of Recent terrestrial mammals represents a roughly stable equilibrium, then for each cladogenesis event in the simulation there must be one extinction event, on average. (This equilibrium assumption is not central to our results, and can be relaxed without impacting the fundamental nature of the model, so long as the total number of extant species grows slowly relative to the rate of species turnover.) Thus, the baseline extinction rate is simply $\beta=1/ n$, where $n$ is the expected number of species at equilibrium. We let $n=5000$, although its precise value is unimportant. By letting extinction rate increase with body size, the actual number of species at equilibrium $n_{\rm eq}$ will be somewhat less than this number. If the true number of terrestrial mammal species is substantially greater than our current estimate of roughly $5000$, or if the assumption of equilibrium is incorrect, then the extinction probability curve can be rescaled by lowering the baseline extinction rate, which does not effect other aspects of the simulation such as the overall shape of the distribution.

We estimate the length of the simulation $t_{\max}$ by estimating the total number of mammalian species since the Cretaceous-Tertiary boundary. We estimate this number as $t_{\max}=\tau\, n/ \nu $, where $\tau$ is the number of years of equilibrium, $\nu$ is the average duration or lifetime of a species, and $n$ is the number of species at equilibrium. We let $\tau\approx60$ My, although its precise value has little impact on the results of the simulation. Estimates of the average duration of a species, however, vary quite widely depending on the data used. In the Alroy data set, \mbox{$\nu=2.32(8)$ My} ($n=1703$; the parenthetical value denotes the standard error in the last digit), while in the NOW data set, \mbox{$\nu=1.52(1)$ My} ($n=14099$). We estimate $\nu$ be the average of these: \mbox{$\nu=1.60(1)$ My,} although its exact value is not important (see Appendix~\ref{appendix:sensitivity} and Fig.~\ref{fig:sensitivity}).

Finally, we estimate the value of $\rho$ by numerically minimizing the distributional distance (see Appendix~\ref{appendix:model:scoring}) between the model and the empirical data for terrestrial mammals (Fig.~\ref{fig:sensitivity1}A). In general, we report results for the power-law model of extinction risk; the fitted value of $\rho$ in the logarithmic model is such that the two risk curves are almost identical (see Fig.~\ref{fig:sensitivity1}B), indicating that the functional form is not important -- both models result in a close-to-linear increase in extinction risk with log-size such that the risk of extinction at each step for the largest species is $56-58\%$ larger than the basal extinction risk ($32-34\%$ for $F(\lambda)$ with log-normal tails). When spread over six or seven orders of magnitude, this causes a slight, positive dependence of extinction risk on body size. We note that the form of this curve provides a testable prediction of the model.

\begin{table}
\begin{center}
\begin{tabular}{l|cccc|c} 
parameter & & & value & & source Ref.\\
\hline
lower bound & & $x_{\min}$  & $1.8$g & & \cite{alroy:2008,smith:etal:2003,fortelius:2003} \\
founder body size & & $x_{0}$  & $40$g & & \cite{smith:etal:2003} \\ 
species at equilibrium & & $n$  & $5000$ & & \cite{smith:etal:2003} \\ 
baseline extinction rate & & $\beta$  & $1/n$ & & --\\ 
rate of extinction increase & & $\rho$  & $0.025$ & & -- \\ 
mean species lifetime & & $\nu$  & $1.60(1)$ My & & \cite{alroy:2008,fortelius:2003} \\ 
years in equilibrium & & $\tau$  & $60$ My & & \cite{alroy:2008} \\ 
$\log \lambda$-intercept & & $c_{1}$  & $0.33$ & & \cite{alroy:2008} \\ 
$\log x$-intercept & & $c_{2}$  & $1.30$ & & \cite{alroy:2008} \\ 
systematic bias & & $\delta$  & $0.04$ & & \cite{alroy:2008} \\ 
variance & & $\sigma$  & $0.63$ & & \cite{alroy:2008} \\ 
power-law tail & & $\alpha$  & $3.3(1)$ & & \cite{alroy:2008} \\
\end{tabular}
\end{center}
\caption{Cladogenetic simulation parameters, their estimated values and the data sources from which the estimates were derived. The parameters can be grouped according to mechanism: the physiological lower limit of the terrestrial mammalian body size ($x_{\min}$); the distribution $F(\lambda)$ of within-lineage changes to body size ($c_{1}$, $c_{2}$, $\delta$, $\sigma$ and $\alpha$), where $\delta$ denotes the systematic bias away from smaller body sizes (Cope's rule) and $c_{1}$ and $c_{2}$ denote the additional bias for small-bodied species; the initial conditions and duration of the simulation ($x_{0}$, $\tau$, $\nu$ and $n$).}
\label{table:params}
\end{table}

\subsection{Scoring the quality of the model}
\label{appendix:model:scoring}
The output of the simulation is a set of species body sizes. To evaluate the quality of this set relative to the empirical data on terrestrial mammals, we use a distance measure for statistical distributions, the tail-weighted Kolmogorov-Smirnov (wKS) goodness-of-fit statistic~\cite{press:etal:1992}
\begin{equation}
  {\rm wKS} = \max_{x} \frac{\left| S(x) - P(x) \right|} {\sqrt{P(x)[1-P(x)]} } \enspace ,
\label{eq:ks}
\end{equation}
where $S(x)$ is the cumulative distribution function (CDF) of the simulated data and $P(x)$ is the CDF of the empirical data. This statistic is independent of any particular binning scheme and thus gives a relatively general characterization of the dissimilarity of two distributions by measuring the maximum absolute deviation between the simulated and empirical cumulative distributions. Very small values (wKS $<0.3$) indicate a strong closeness, for all values of $x$. In Fig.~\ref{fig:appendix:model}C, for instance, \mbox{wKS $\approx 0.17$}.

Some readers may be familiar with the more commonly used Kolmogorov-Smirnov (KS) goodness-of-fit statistic. The tail-weighted version differs by giving equal weight to all parts of the distribution, and particularly the tails. In contrast, the traditional KS statistic effectively weights the area near the median of the distribution the most, and thus can underestimate strong differences in the tails. This causes the tail-weighted version to be more difficult to minimize -- all parts of the simulated distribution must be close to the empirical one, not just the middles of the distributions. We have tried using both statistics to score the quality of the model results, and we find that numerically minimizing the tail-weighted version chooses values of $\rho$ that produce significantly more convincing results for larger-bodied species, e.g., $x>10^{4}\g$.

Finally, because the model produces a dynamic equilibrium in the species body size distribution, to evaluate its typical behavior, and to prevent transient effects from skewing our quality scores, we average the wKS statistic over regularly spaced intervals in the last 15 My of simulated time. When we evaluate the quality of a set of parameter values, we further average this value over several hundred independent trials.

\section{Changes to body size and Cope's rule}
\label{appendix:copesrule}
Rule 2 represents the manner in which body sizes vary at speciation events. Phylogenetic body size data for a wide range of terrestrial mammals would be the preferred way to determine the best model of within-lineage body size variation, but such ancestor-descendent data is not currently available for a sufficiently large and diverse set of terrestrial mammals. Instead, we use Alroy's putative ancestor-descendant data, reconstructed from fossil data for North American mammals, as a proxy. This data has been used in several previous studies of within-lineage variation of body size~\cite{alroy:2008,alroy:1998}, and details of the non-phylogenetic reconstruction process for the 1106 pairs of terrestrial mammals species are given there. From this data, we estimate a parametric model for $F(\lambda)$.

The non-phylogenetic nature of this data, however, implies that there are likely to be several inversions of ancestors and descendants, as well as several incorrect pairings of ancestors with descendants. Fortunately, the statistical nature of our analysis implies that so long as the number of putative pairs is relatively large, such errors will not obscure the true average log-change, which is precisely the aspect of this data most important to our study. Further, our sensitivity analysis indicates that the precise details of the inferred model, e.g., the average and variance, do not matter much with regard to our overall conclusions (see Appendix~\ref{appendix:sensitivity}), so long as a log-normal model of change is a relatively good model of the data.

\subsection{Empirical evidence for Cope's rule}
\label{appendix:copesrule:evidence}
Empirical evidence for and against Cope's rule has been studied in a variety different taxonomic groups~\cite{macfadden:1986,jablonski:1997,maurer:1998,alroy:1998,bokma:2002,valkenburgh:etal:2004,novackgottshall:lanier:2008}. For terrestrial mammals, the evidence is relatively strong, with Alroy's study~\cite{alroy:1998} showing a slight systematic positive bias $\langle\log\lambda\rangle>0$, with descendants tending to be slightly larger than their ancestors.

\begin{figure}[t]
\begin{center}
\includegraphics[scale=0.45]{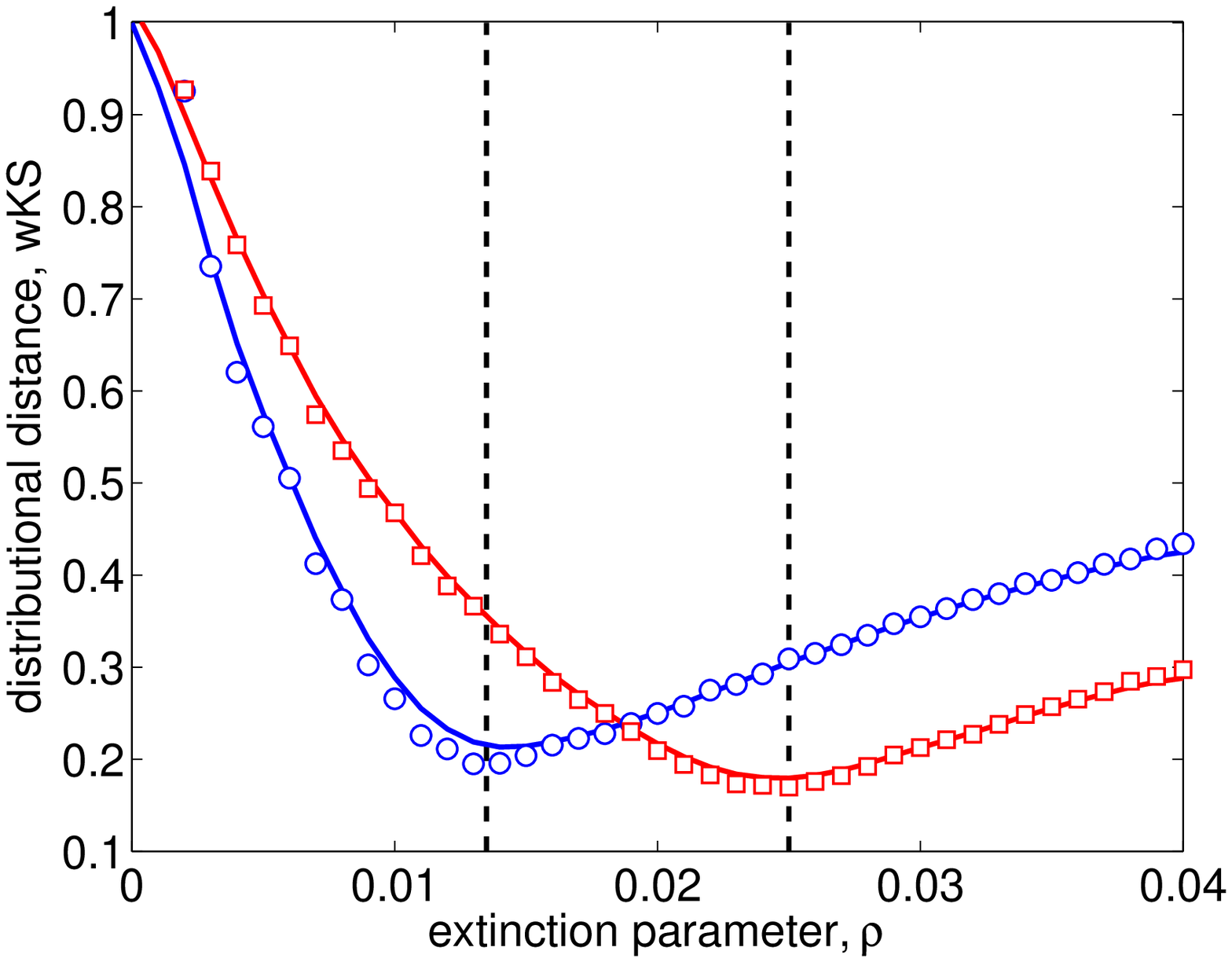} \\
\includegraphics[scale=0.45]{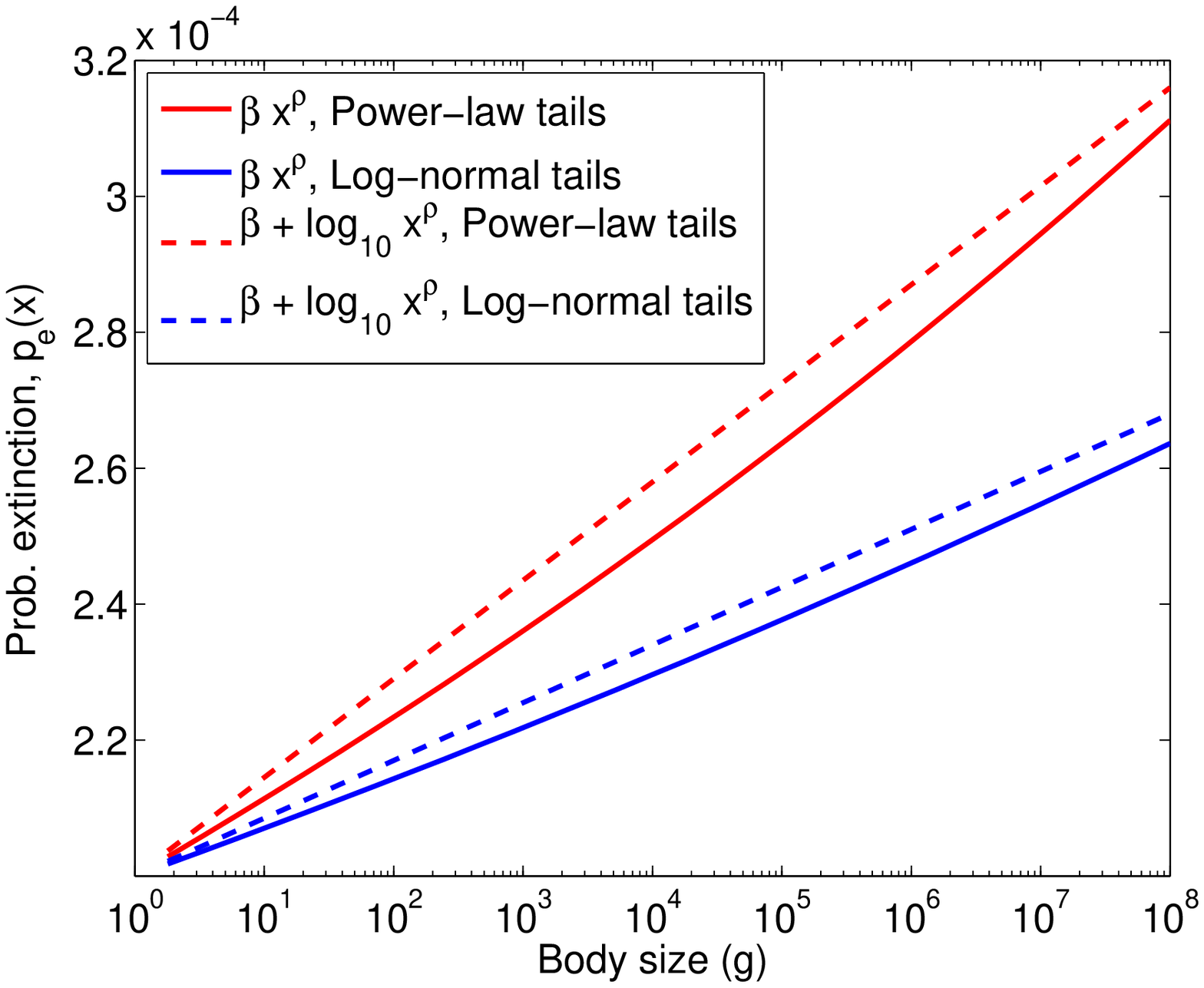} 
\end{center}
\caption{(\textbf{A}) Estimation results for fitting the free parameter $\rho$ in the power-law model of extinction risk, in two alternative cases, one where the distribution $F(\lambda)$ of within-lineage changes to body size has log-normal tails (blue), and one where the tails decay as a power (red). Similar results are obtained under the logarithmic model of extinction risk. All other parameters take the values given in Table~\ref{table:params}. For clarity, we also plot a smoothed trend (exponential kernel) over the sampled data. Each point is the average goodness-of-fit $\langle$wKS$\rangle$, for the last $15$ My of the simulation, over $50$ independent trials. (\textbf{B}) The fitted extinction-risk curves for models of $F(\lambda)$ with power-law and non-power-law tails, and for models where the extinction risk increases as a logarithm or power of size (see Appendix~\ref{appendix:model:specification}, Rule 3). The similarity of the curves between these two extinction models shows that a generally log-linear form is sufficient.}
\label{fig:sensitivity1}
\end{figure}

In order to specify Rule 2, however, we need to know not only whether there is a positive bias or not, but how strong is the bias as a function of ancestor size. This can be done by directly estimating the shape of $F(\lambda)$ as a function of ancestor size. Thus, we conduct a new analysis of the previously studied ancestor-descendant data.

\begin{figure*}[t]
\begin{center}
\begin{tabular}{cccc}
\includegraphics[scale=0.233]{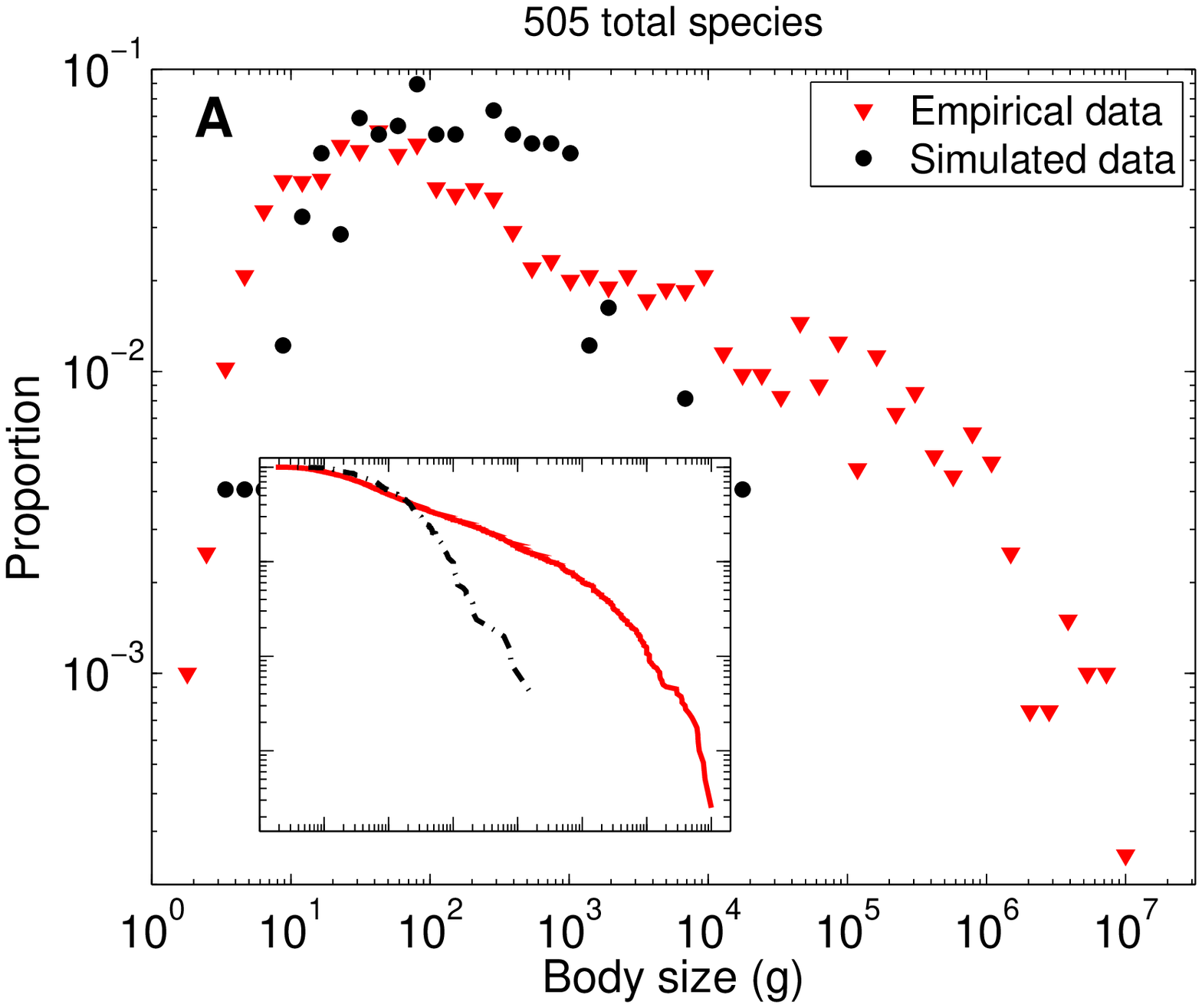} & 
\includegraphics[scale=0.233]{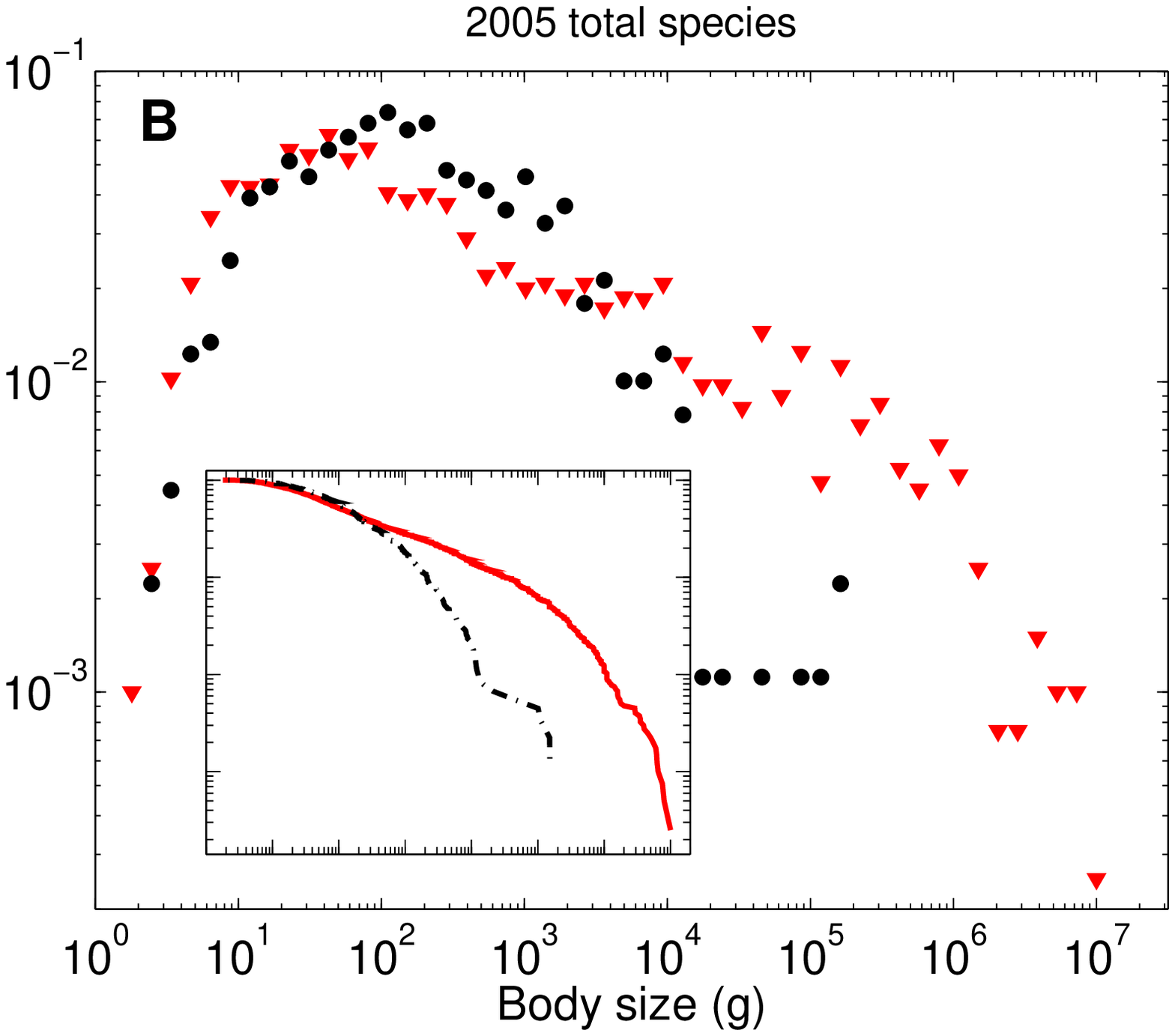} & 
\includegraphics[scale=0.233]{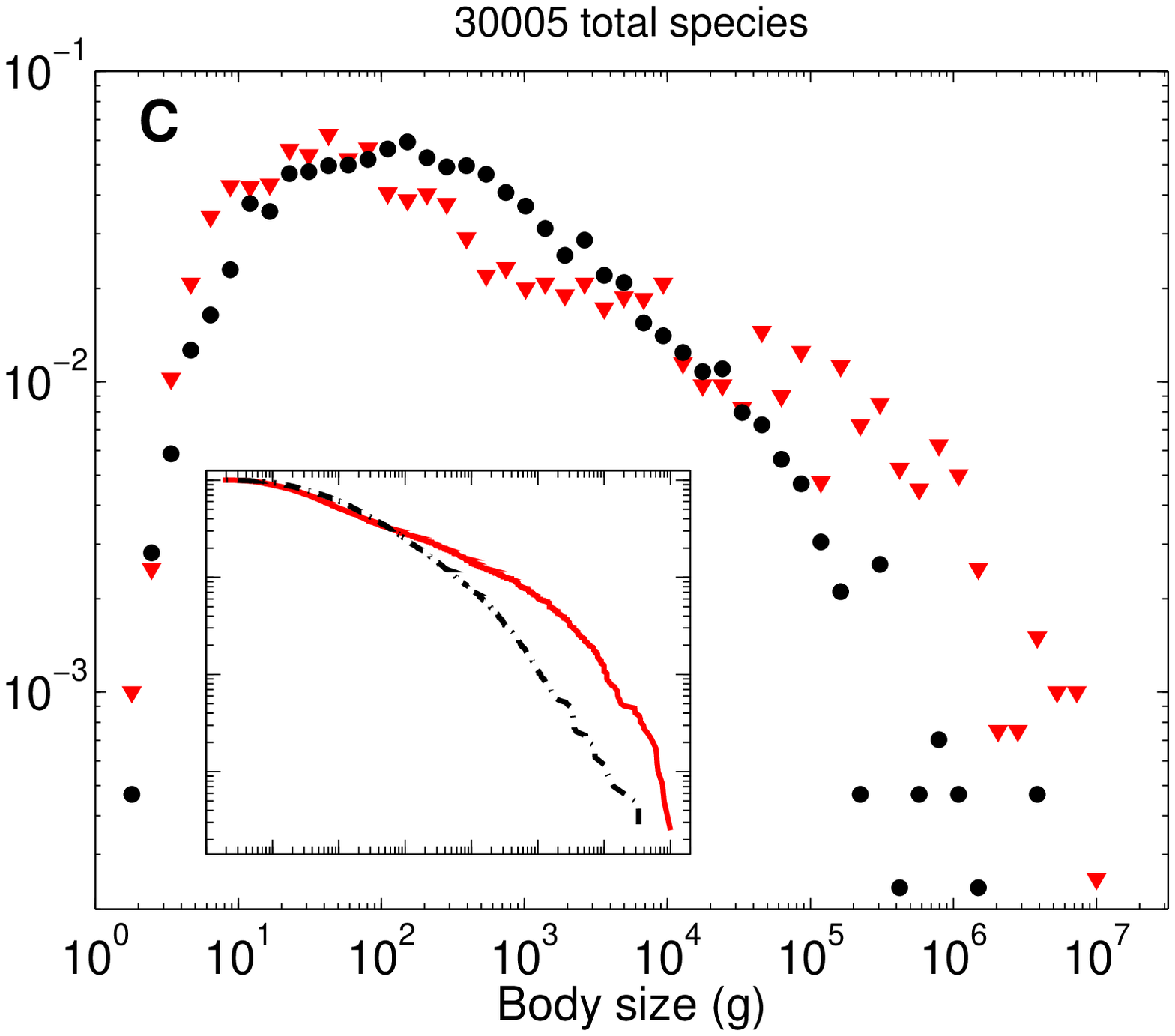} &
\includegraphics[scale=0.233]{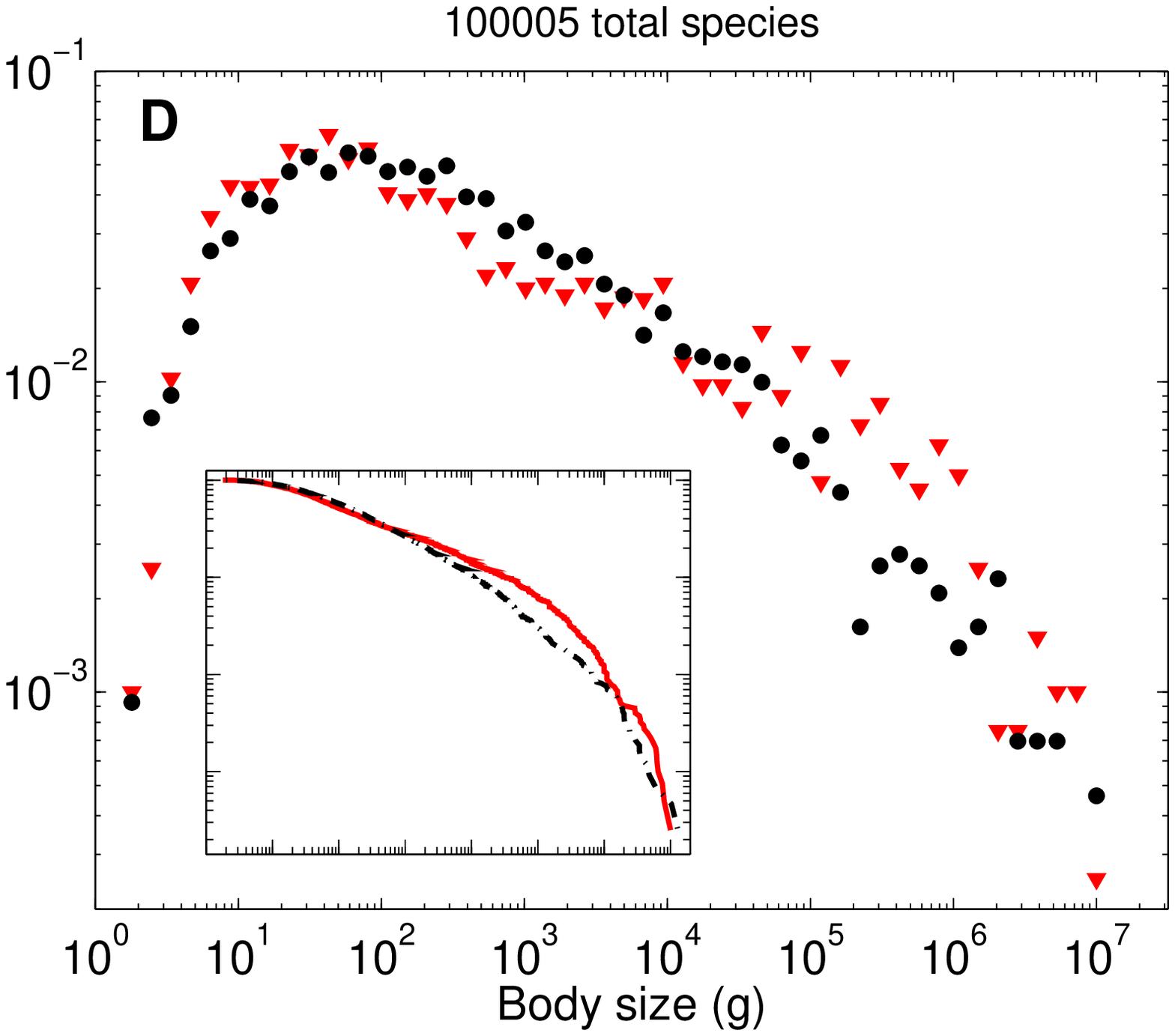} 
\end{tabular}
\end{center}
\caption{Snapshots of the simulated species body size distribution, relative to the empirical distribution, from a single simulation trial, taken at $n=\{505,2005,30005,100005\}$ total species (\textbf{A}, \textbf{B}, \textbf{C} and \textbf{D}, respectively). For clarity, the insets show the corresponding complementary cumulative distribution functions.}
\label{fig:timeseries}
\end{figure*}
\begin{figure}[t]
\begin{center}
\includegraphics[scale=0.45]{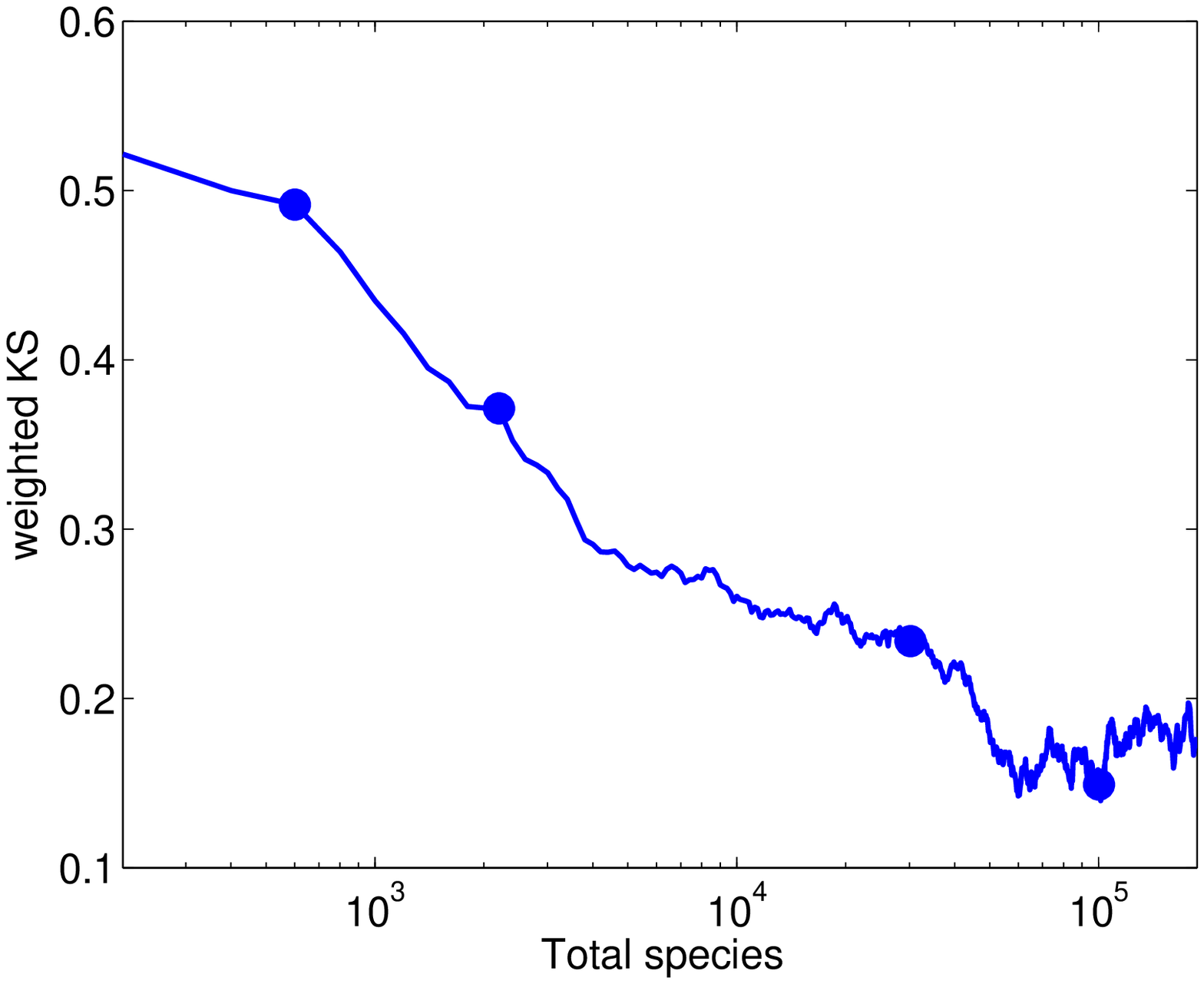} 
\end{center}
\caption{The time series of wKS statistics for the simulation in Fig.~\ref{fig:timeseries}. The bold circles indicate the positions and scores of the four snapshots.}
\label{fig:timeseries2}
\end{figure}

Fig.~\ref{fig:paleo}A shows descendant body size $x_{D}$ as a function of ancestor body size $x_{A}$, for Alroy's fossil data on North American mammals, and illustrates that descendants tend to be roughly the same size as their ancestors. The best-fit allometric relation~\cite{warton:etal:2006} $\log x_{D}=\tilde{\lambda}\log x_{A}$ to these data yields $\tilde{\lambda}=1.02\pm0.01$ (estimate $\pm 95\%$ confidence), indicating a small but systematic tendency for descendants to be slightly larger than their ancestors.

Fig.~\ref{fig:paleo}B shows the distribution of within-lineage changes in body size (equivalent to the vertical residuals to the line $x_{D}=x_{A}$ in Fig.~\ref{fig:paleo}A), with increases ($615$) being only slightly more common than decreases ($488$; the remaining $3$ cases are instances of no-change). Denoting $\lambda$ as the multiplicative change in body size from ancestor to descendant, we find that the overall average change is toward larger sizes, with \mbox{$\langle\log\lambda \rangle=0.047\pm0.009$}. This estimate ignores, of course, the possibility that the average change depends on the ancestor size.

The conventional assumption in simulation studies of body size evolution is that $F(\lambda)$ follows a log-normal distribution. We find that the data are consistent with this assumption; however, we note that the data are also consistent with a log-normal double Pareto distribution~\cite{reed:jorgensen:2004} -- a log-normal distribution with tails that decay as power-laws (or, that decay as exponentials in $\log \lambda$). We test this hypothesis using standard statistical techniques for power-law distributions~\cite{clauset:etal:2007}, and find that the tails of the distribution can be assumed to be symmetric [negative tail: $\alpha=3.4(2)$, $p=0.83(3)$; positive tail: $\alpha=3.3(2)$, $p=0.79(3)$; both tails together: $\alpha=3.3(1)$, $p=0.96(3)$]. For completeness, we consider both models of $F(\lambda)$ in our sensitivity analysis, and find relatively small differences between the results (but see Appendix~\ref{appendix:alternatives}).  

\begin{figure*}[t]
\begin{center}
\includegraphics[scale=0.311]{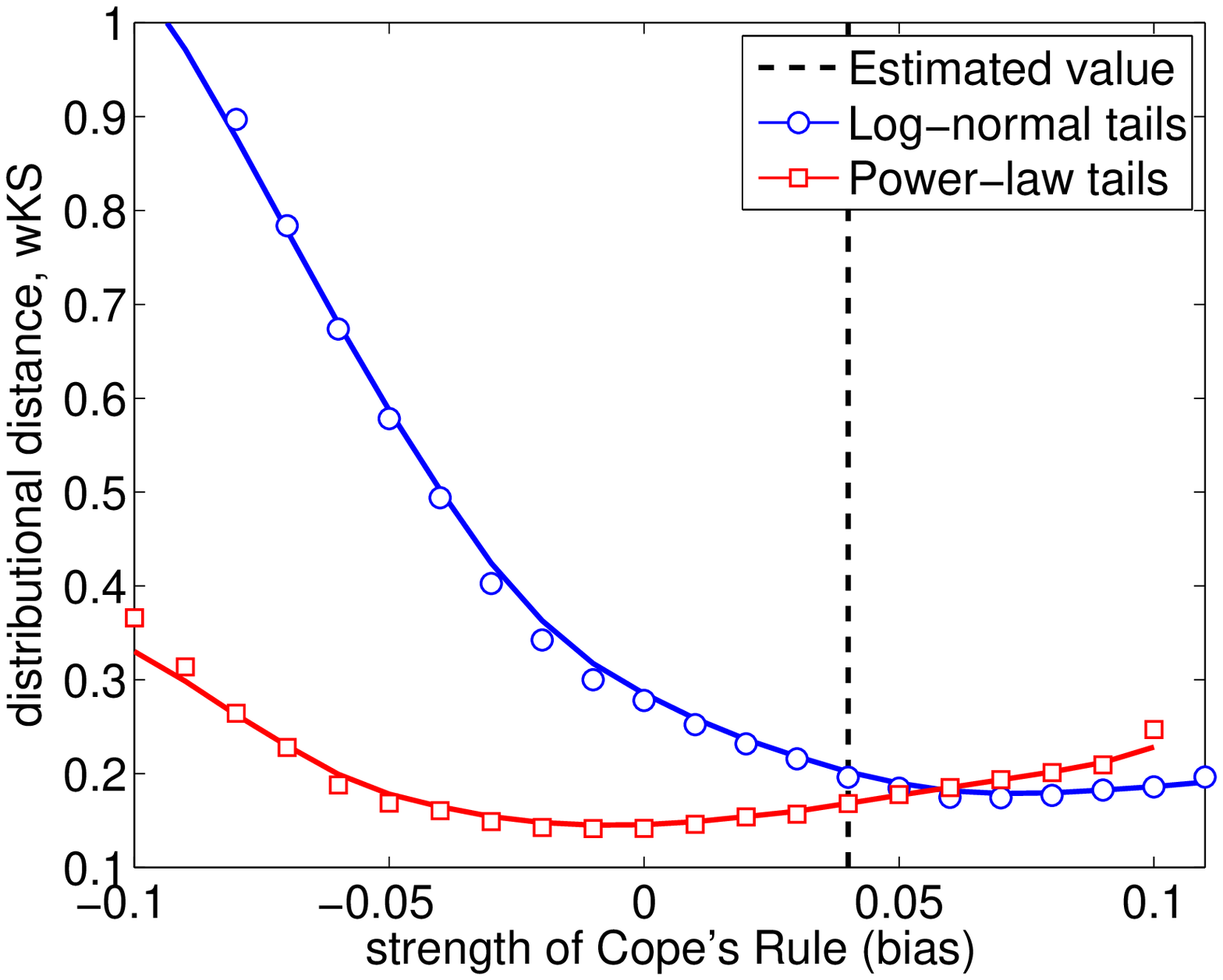} 
\includegraphics[scale=0.311]{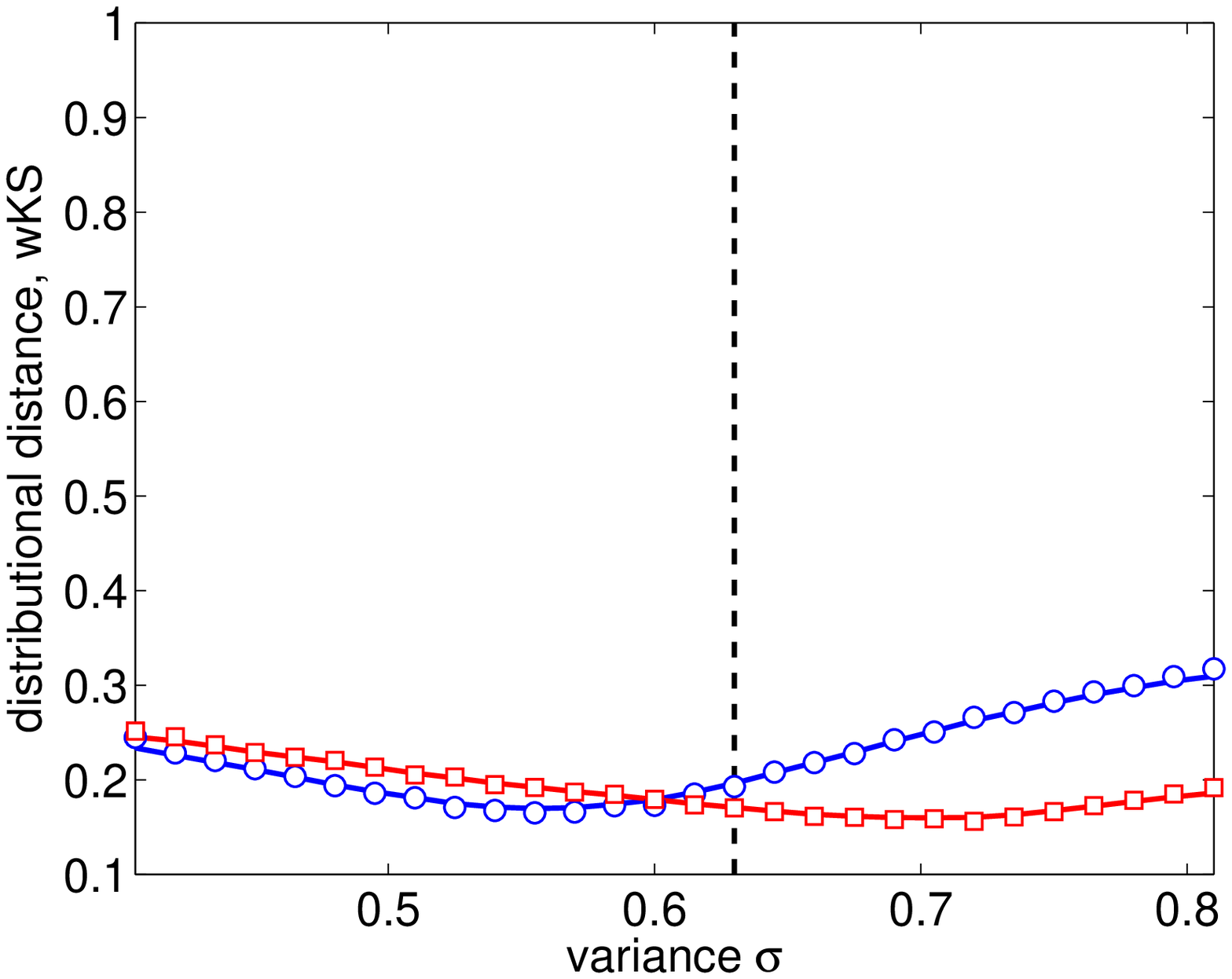} 
\includegraphics[scale=0.311]{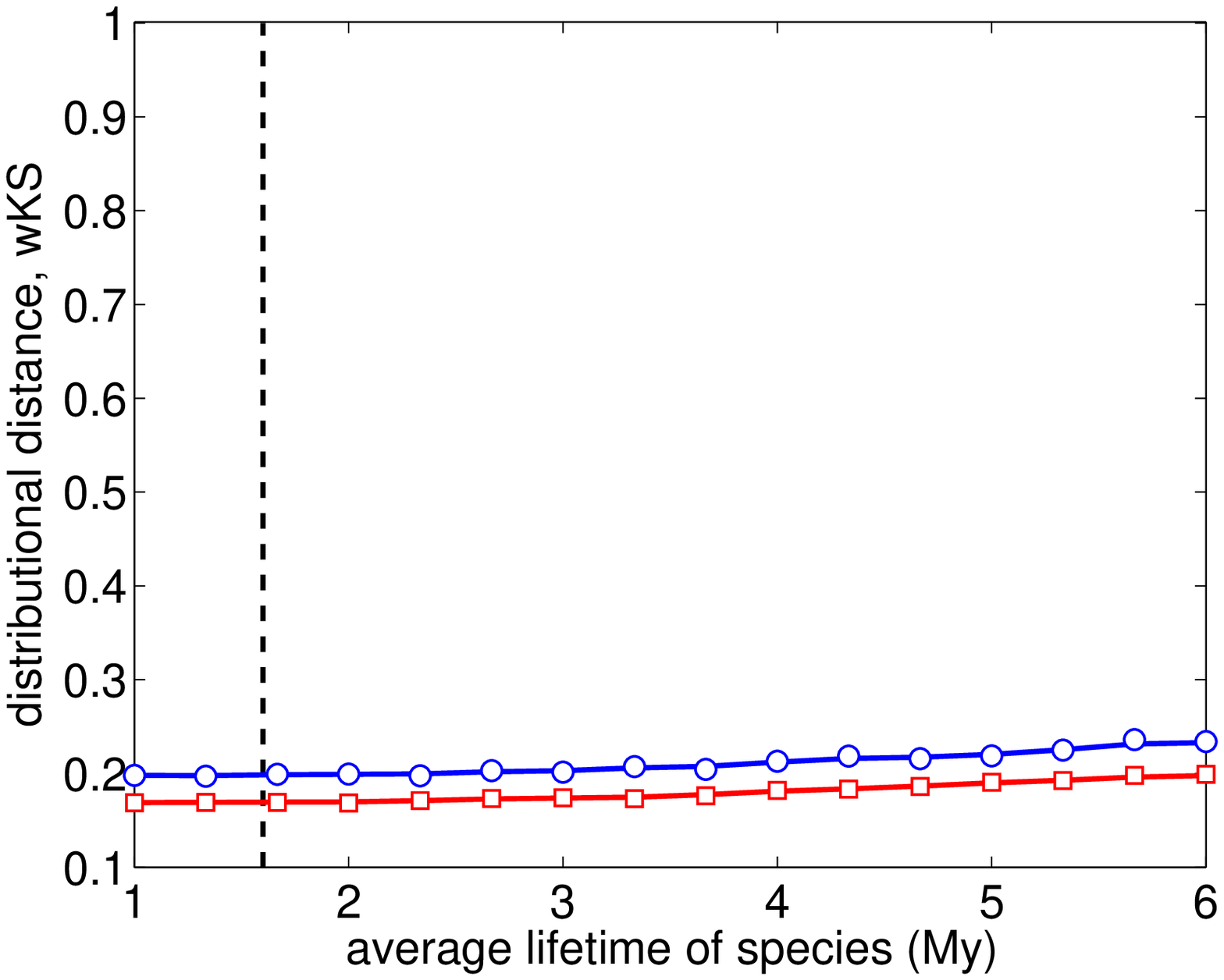} 
\includegraphics[scale=0.311]{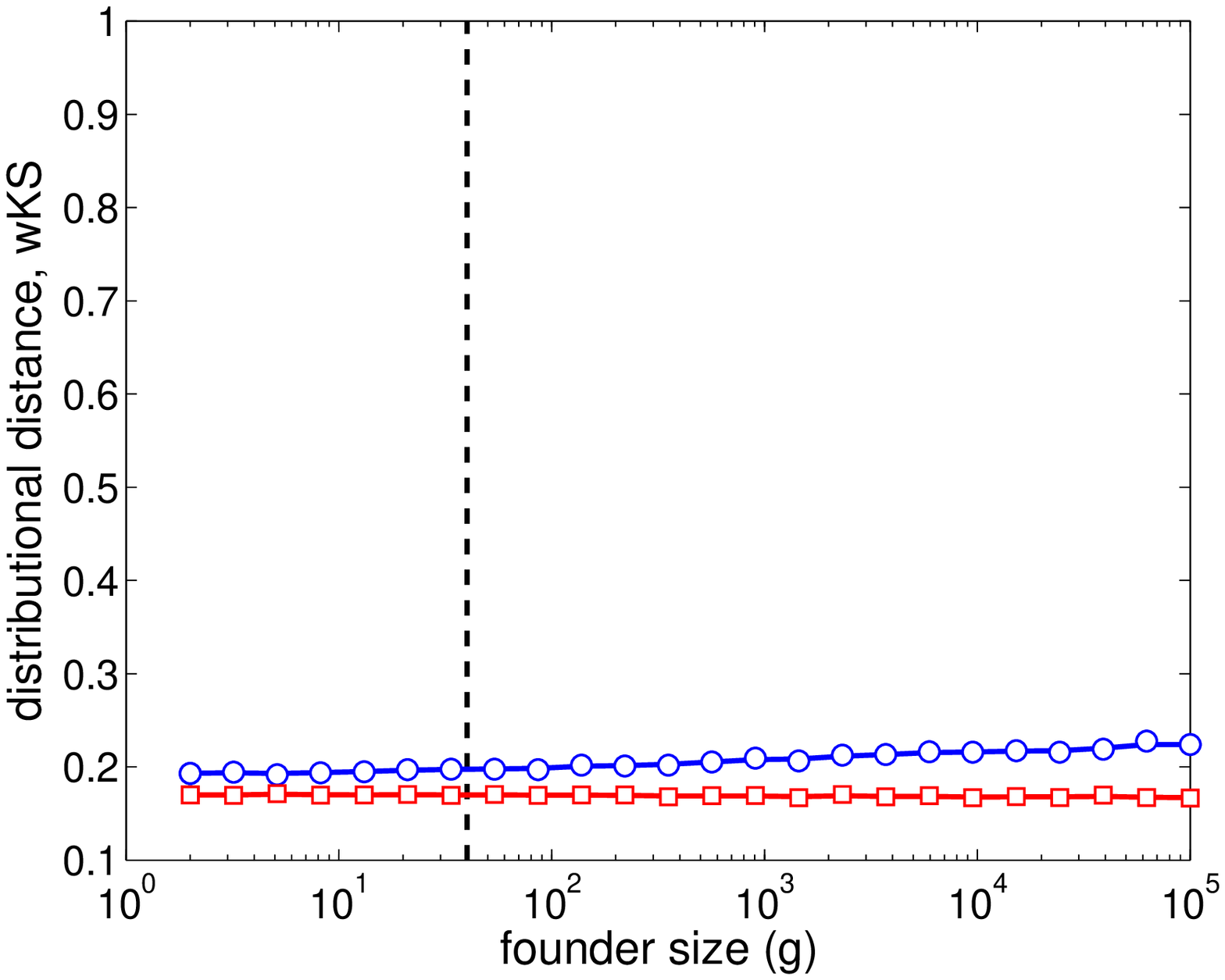} 
\includegraphics[scale=0.311]{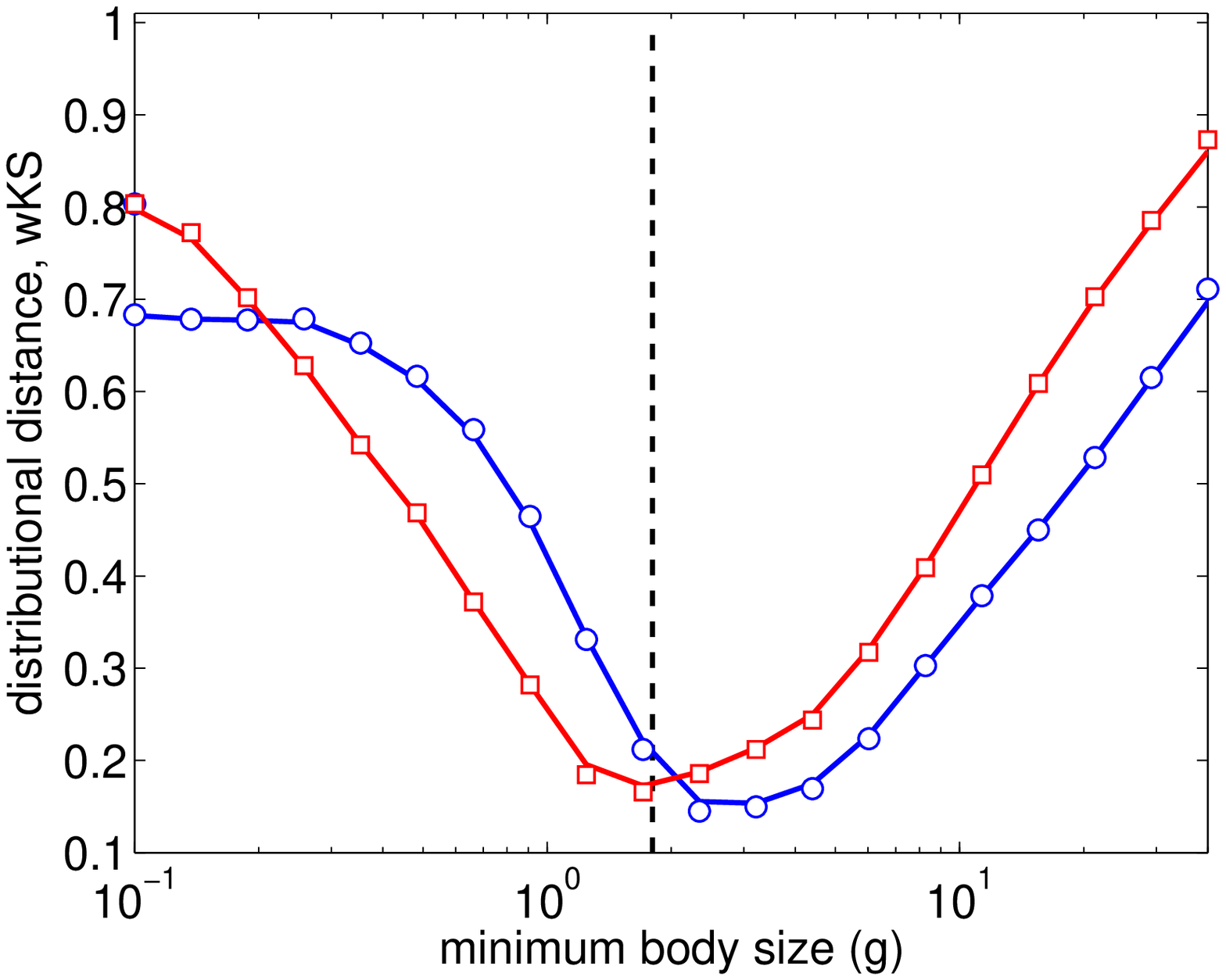} 
\end{center}
\caption{Sensitivity analysis of the quality of the simulated species body size to variations in the values of the model parameters estimated from data (Table~\ref{table:params}). Each figure shows the results for the model $F(\lambda)$ with (red squares) and without (blue circles) power-law tails; for clarity, we also plot a smoothed trend (exponential kernel) over the results. Each point denotes the $\langle$wKS$\rangle$ statistic, averaged over the last $15$ My of the simulation and over $100$ independent trials. Further, because $\rho$ is a free parameter of the model, for each point, we re-estimated $\rho$ as the value that gave the minimum $\langle$wKS$\rangle$ (over $100$ independent trials), given the choice of the parameter in question, with all other parameters being held fixed.}
\label{fig:sensitivity}
\end{figure*}

\subsection{Our model of changes to body size}
\label{appendix:copesrule:model}
In this section, we describe a model-selection analysis among three alternative models of within-lineage changes to body size $F(\lambda)$, all of which are drawn from a log-normal distribution where the average log-change to size $\mu$ depends on the ancestor's size $x_{A}$. In this way, $F(\lambda)$ can model both the effect of Cope's rule on large-bodied species \emph{and} the effects of constrained evolution near the lower limit of body size on real mammalian evolution (above and beyond the form imposed by respecting the lower limit in Rule 2). This latter effect we call the small-bodied bias. For these three models, we ask which has the best empirical support from the putative ancestor-descendent data.

\begin{enumerate}
\item Model one is a piece-wise form in which a bias toward larger sizes for small-bodied species decreases as a power of body size to a constant value $\delta$ for large-bodied species (Fig.~\ref{fig:paleo}C).
\item Model two is identical to model one but sets the large-body bias parameter $\delta$ to zero. 
\item Model three is a function $\mu$ that follows the best-fit cubic polynomial (see~\cite{alroy:1998}).
\end{enumerate}
All models have the form \mbox{$\log \lambda(x_{A})=\mathcal{N}[\mu(x_{A}),\sigma^{2}]$} -- that is, $\log\lambda$ is normally distributed with constant variance $\sigma$ and a mean $\mu$ that varies as a function of body size $x_{A}$, where the particular functional form of $\mu(x_{A})$ varies from model to model. In the first two cases, we use a simple piece-wise linear function:
\begin{equation}
\label{eq:muxa}
\mu(x_{A}) = \left\{\begin{array}{ll}
           (c_{1} / c_{2}) \log x_{A} + c_{1} + \delta & \quad\mbox{if $\log x_{A} <  c_{2}$,} \\
           \delta & \quad\mbox{otherwise,}
         \end{array}
         \right.
\end{equation}
where $c_{1}$ is the $y$-intercept and $c_{2}$ is the $x$-intercept of the size-dependent bias for small bodies, and $\delta$ is the magnitude of the systematic positive bias for larger species. Thus, $c_{1}$ controls the strength of the small-body bias and $c_{2}$ controls the range over which this bias decays; their ratio $-c_{1}/c_{2}$ gives the power by which the bias toward larger descendants decreases with increasing ancestor size. When \mbox{$\delta=0$} (model two), there is no systematic bias toward larger bodied species; a bias toward larger descendants (Cope's rule) is modeled by \mbox{$\delta>0$} (model one). In the third case, we let $\mu(x_{A})$ be the best-fit third-order polynomial to the ancestor-descendent data~\cite{alroy:1998}; this function crosses the $x$-axis in three places, implying the existence of two ``optimal'' body sizes, one for small-bodied and one for large-bodied species. In all cases, we estimate the free parameters of these models from the data using maximum likelihood.

Although each of the three models fits the data reasonably well (\mbox{$p>0.1$} under a standard parametric bootstrap test~\cite{efron:tibshirani:1993}), the data is closest to model one [likelihood ratio test (LRT)~\cite{vuong:1989}, \mbox{$| \log (L_{1} / L_{2})|=7.226$}, \mbox{$p=1.44\times 10^{-4}$} and \mbox{$| \log (L_{1} / L_{3})|=1.001$}, \mbox{$p=0.84$}, with similar results for a Bayesian Information Criterion (BIC) comparison].

Fig.~\ref{fig:paleo}C shows the fitted form of the best model, where the strength of Cope's rule is \mbox{$\delta=0.04\pm0.01$} for \mbox{$x_{A}\gtrsim 32\g$} (or, an average growth of $4.1\pm1.0\%$ per speciation event), along with the raw ancestor-descendent data. This model is visually very similar to a smoothed version of the data~\cite{wasserman:2006}, shown in Fig.~\ref{fig:smoothed}. Although these results suggest that our estimated model is a good summary of the data, the data themselves could be biased in several ways. A more robust analysis would combine the likelihood ratio test approach employed here with an appropriate model of the errors and bias, were such an error-model known for this kind of data.

Finally, we note that the fitted power-law model of the bias toward larger descendants for small-bodied ancestors has an exponent $\gamma\approx-1/4$, which may or may not be related to the prevalent quarter-power scaling in ecology~\cite{savage:etal:2004}.

\section{Additional model results and analysis}
\label{appendix:model:additional}

\subsection{Simulation results}
\label{appendix:model:results}

\begin{table*}
\begin{center}
\begin{tabular}{l|c|ccccc|c} 
 &  & lower & Cope's & \multicolumn{2}{c}{small-body} & extinction & \\
 &  & boundary & rule & \multicolumn{2}{c}{bias} & dependence & \\
 \hline
Model description & $\langle$wKS$\rangle$ & $x_{\min}$ & $\delta$ & $c_{1}$ & $c_{2}$ & $\rho$ & Fig. \\
\hline
Full model & $0.181(1)$ & $1.8$ & $0.04$ & $0.33$ & $1.30$ & $0.025$ & \ref{fig:comparison:1}A \\
Full model with & \multirow{2}{*}{$0.244(1)$} & \multirow{2}{*}{$1.8$} & \multirow{2}{*}{$0.04$} & \multirow{2}{*}{$0$} & \multirow{2}{*}{$0.25$} & \multirow{2}{*}{$0.023$} & \multirow{2}{*}{\ref{fig:comparison:1}B} \\
\hspace{3mm} no small-size bias & & & & & & & \\
Unbiased diffusion & \multirow{2}{*}{$2.97(3)$} &  \multirow{2}{*}{$1.8$} & \multirow{2}{*}{$0$} & \multirow{2}{*}{$0$} & \multirow{2}{*}{$0.25$} & \multirow{2}{*}{$0$} & \multirow{2}{*}{\ref{fig:comparison:1}C} \\
\hspace{3mm} with lower bound & & & & & & & \\
\hline
Cope's rule with size- & \multirow{2}{*}{$10.60(7)$} & \multirow{2}{*}{$10^{-8}$} & \multirow{2}{*}{$0.04$} & \multirow{2}{*}{$0$} & \multirow{2}{*}{$-8$} & \multirow{2}{*}{$-0.002$} & \multirow{2}{*}{\ref{fig:comparison:2}D} \\
\hspace{3mm} dependent extinction & &  & & & & &  \\
Cope's rule alone & $11.72(9)$ & $10^{-8}$ & $0.04$ & $0$ & $-8$ & $0$ & \ref{fig:comparison:2}E \\
Size-dependent & \multirow{2}{*}{$10.37(6)$} & \multirow{2}{*}{$10^{-8}$} & \multirow{2}{*}{$0$} & \multirow{2}{*}{$0$} & \multirow{2}{*}{$-8$} & \multirow{2}{*}{$-0.005$} & \multirow{2}{*}{\ref{fig:comparison:2}F} \\
\hspace{3mm} extinction alone & & & & & & & 
\end{tabular}
\end{center}
\caption{A comparison of the full model described in Appendix~\ref{appendix:model:specification} with five simpler models. Each of these alternatives are special cases of the full model and many have been discussed in the literature (see~\cite{kozlowski:gawelczyk:2002,allen:etal:2006}) as methods for generating right-skewed size distributions. Each model was run 1000 times, from which we computed the central tendency of the simulated distribution (shown in Figs.~\ref{fig:comparison:1} and~\ref{fig:comparison:2}) and the average statistical distance $\langle$wKS$\rangle$ from the empirical distribution. Results reported here are for $F(\lambda)$ with power-law tails and the power-law model of extinction risk (similar results for log-normal tails or logarithmic extinction risk); the standard error in the last digit is quoted parenthetically. For models with $\rho\not=0$, $\rho$ was estimated by minimizing $\langle$wKS$\rangle$.} 
\label{table:comparison}
\end{table*}

To convey some notion of how the simulation develops the species body size distribution over time, Fig.~\ref{fig:timeseries} shows snapshots of simulated data, along with the empirical data, taken from a single run of the simulation. Initially, the simulated distribution is concentrated around the size of the founder species $x_{0}$, but, over time, the distribution's right tail lengthens considerably until the simulated distribution is very close to the empirical one, for all body sizes. After approximately 30~000 total simulated species (Fig.~\ref{fig:timeseries}C), the agreement between the simulation and data is already relatively good (wKS $=0.37$), with the main disagreement being for the largest-bodied species. By this point, the disagreement for small- and intermediate-bodied species is very small.
Fig.~\ref{fig:timeseries2} shows the corresponding time series of the wKS statistic over simulated time, for the same simulation.

\subsection{Sensitivity analysis}
\label{appendix:sensitivity}
We tested the dependence of our results on the particular estimated parameter values by conducting a thorough sensitivity analysis that varied each parameter independently over a wide range of values. For each of these alternative parameterizations, we re-estimated the value of the free parameter $\rho$ (by repeating the calculation shown in Fig.~\ref{fig:sensitivity1}A). \mbox{Fig.~\ref{fig:sensitivity}} shows the results of these tests, for two models of $F(\lambda)$, one with log-normal and one with power-law tails. Results from the logarithmic model extinction risk are omitted as they are virtually indistinguishable from the results from the power-law model.

Typically, the precision of the simulated distribution is highly robust to variations in the estimated values of most model parameters, with \mbox{wKS $<0.3$} and deviations appearing only in the extreme tails or in the $1\kg$ or $300\kg$ ranges. In particular, the precision is highly insensitive to the size of the founder species $x_{0}$ or the length of the simulation (parameterized by the average lifetime of a species $\tau$), and only mildly sensitive to the variance in the diffusion process $\sigma$. Somewhat greater sensitivity is seen for the strength of Cope's rule $\delta$, although both positive and negative values both produce good fits to the data. The most sensitive parameter is the value of the lower limit $x_{min}$, with good fits only being produced when $x_{\min}\approx2\g$.

\begin{figure*}[t]
\begin{center}
\begin{tabular}{cc}
\includegraphics[scale=0.41]{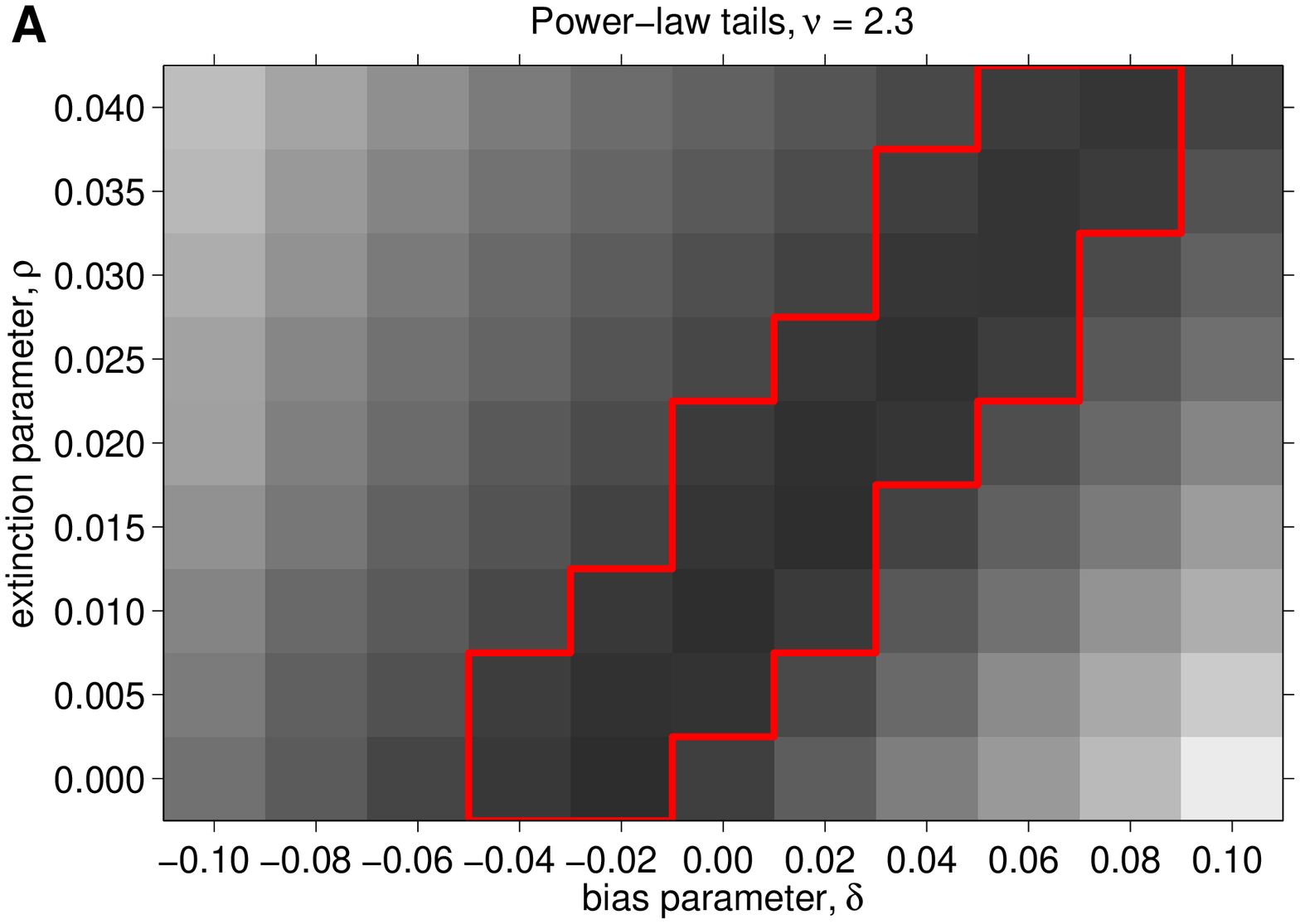} & 
\includegraphics[scale=0.41]{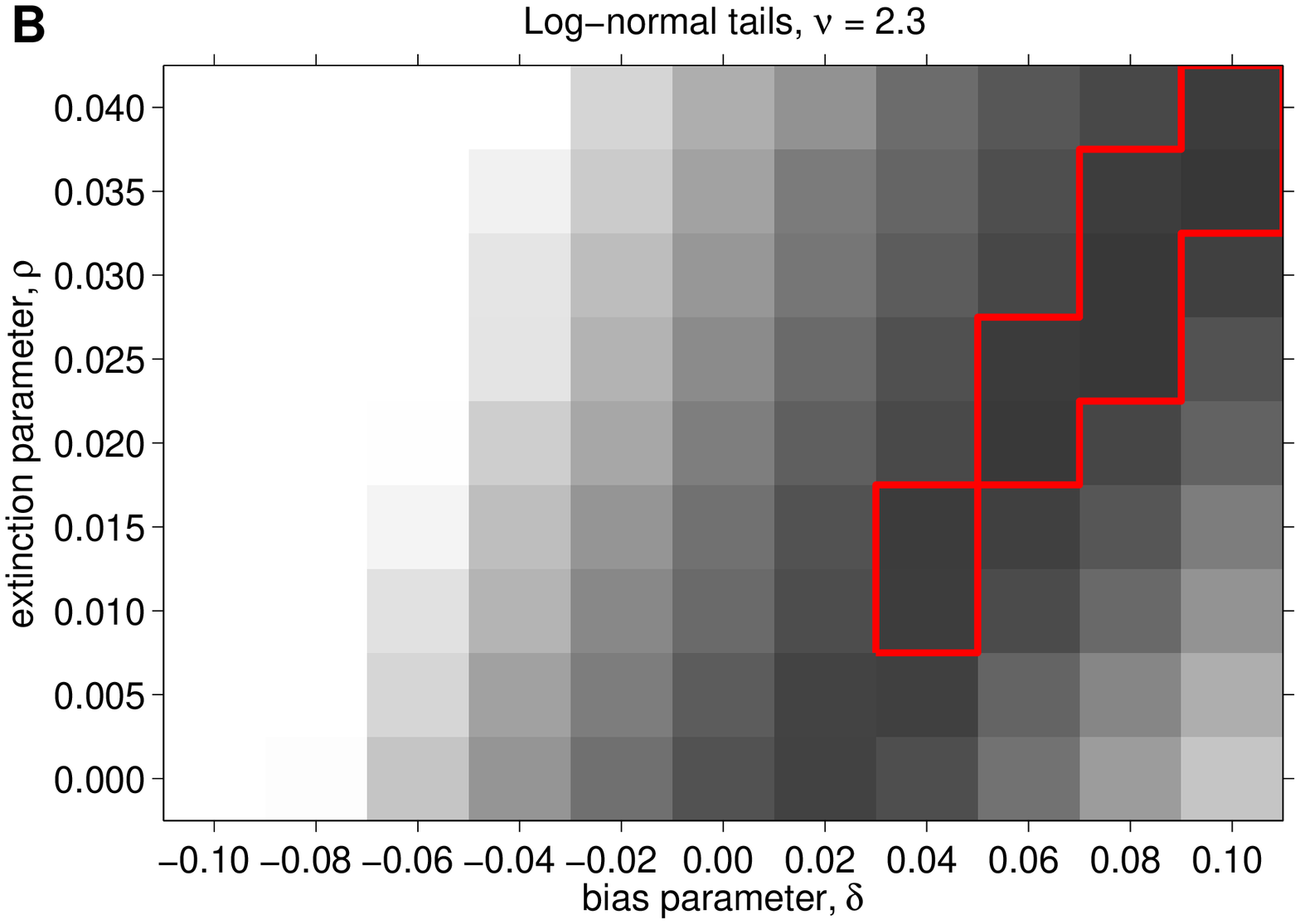} \\
\includegraphics[scale=0.41]{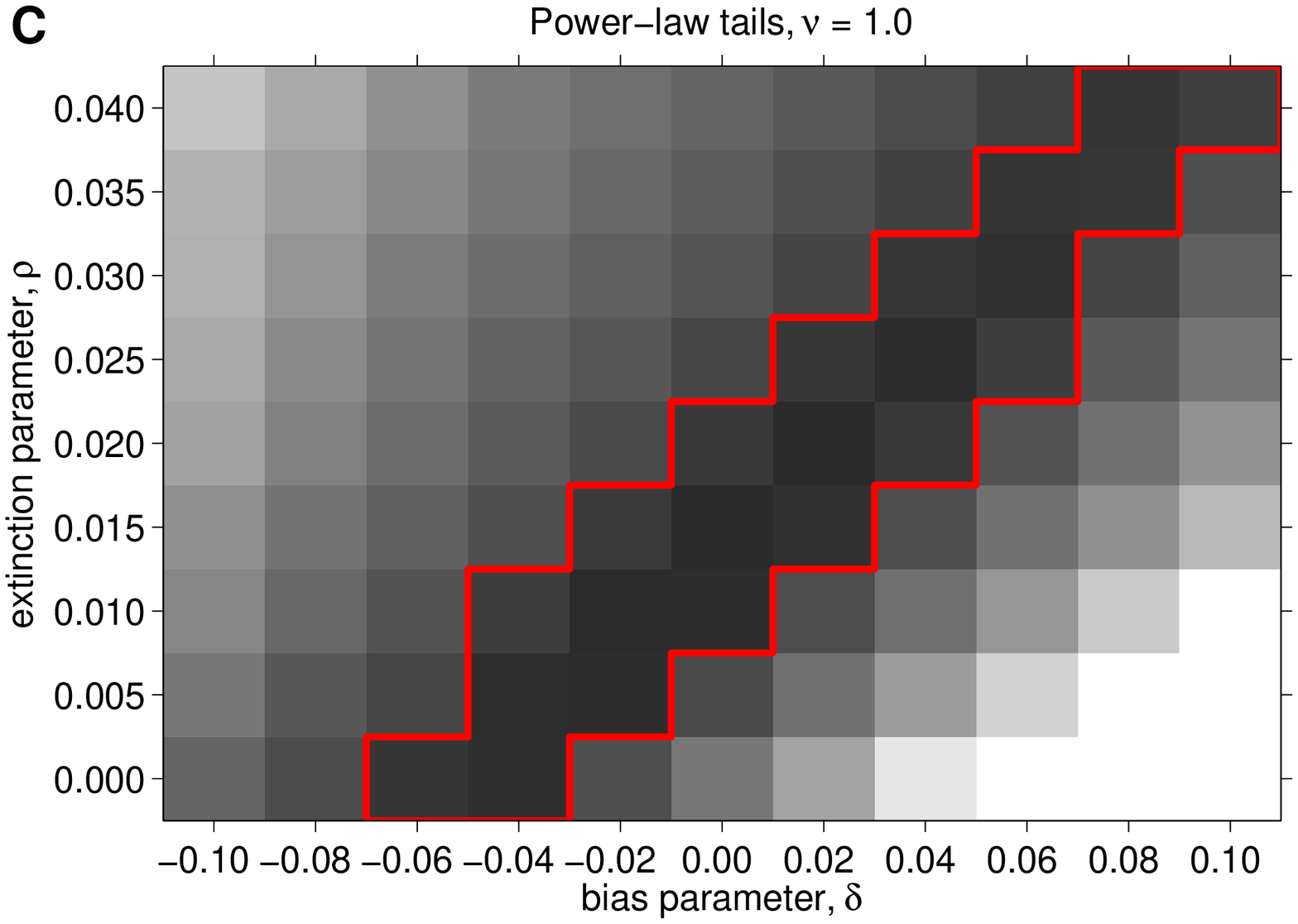} &
\includegraphics[scale=0.41]{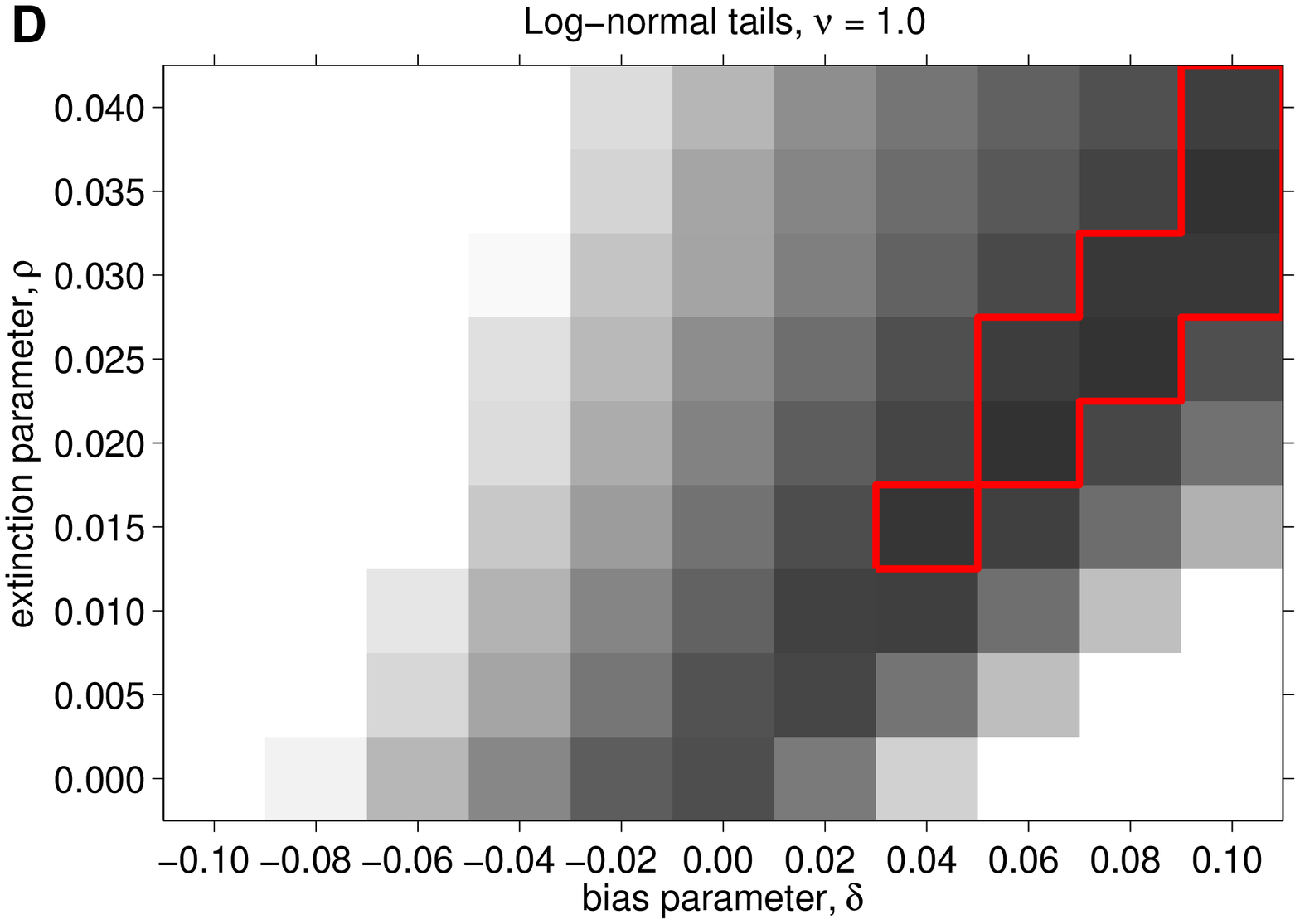} 
\end{tabular}
\end{center}
\caption{Sensitivity analysis for power-law tails in the distribution of body-size changes and the average lifetime of species $\nu$. In each case, we systematically varied both the strength of Cope's rule $\delta$ and the strength of extinction for larger body sizes $\rho$, and computed the average goodness-of-fit to the empirical distribution function, for the last $15$ My of the simulation, over $100$ independent trials. In each figure, we circle the region of parameter space that provides the best fit to the data, \mbox{$\langle$wKS$\rangle$ $\leq 0.25$}. (\textbf{A}, \textbf{C}) show results for using a log-normal distribution with power-law tails (also known as a log-normal double Pareto); (\textbf{B}, \textbf{D}) show results for the same log-normal distribution but without power-law tail. (\textbf{A}, \textbf{B}) show results for \mbox{$\nu=2.3$ My}, while (\textbf{C}, \textbf{D}) show results for \mbox{$\nu=1.0$ My}. Notably, the model with log-normal tails has a much more narrow range of parameter values that provide good fits to the data. For the models with power-law tails, an extinction parameter of \mbox{$0.02 \leq \rho \leq 0.03$} provides the best fit to the data for the particular strength of Cope's rule we estimated from fossil data \mbox{$\delta=0.04$}, regardless of how long the simulation is run.}
\label{fig:systematic}
\end{figure*}

We performed a second sensitivity test to probe the connection between the strength of Cope's rule $\delta$ and the rate of increasing risk from extinction for larger bodied species $\rho$. By systematically varying these two parameters, we find that the particular shape of the right-tail of the empirical distribution can only be produced when these two parameters co-vary in a very regular fashion. Fig.~\ref{fig:systematic} shows the results of this experiment, where we choose two different values of $\nu$ (average species lifetime) and two different forms for $F(\lambda)$ (as before, log-normal or power-law tails).

We interpret these results in the following way. The greater the short-term selective benefits derived from increased species body size, the more species tend to have larger body size, at the expense of smaller body size. If the increased risk of extinction from increased body size does not increase in a related way, then the distribution of species body sizes becomes more heavily weighted toward large-bodied species. If, however, the risk of extinction increases proportionally to the increased benefits of body size, then the size distribution's steady-state remains unchanged.

\section{Comparison with simple diffusion models}
\label{appendix:alternatives}
Less complex diffusion models have also been suggested as possible explanations of right-skewed (on a log-scale) species body size distributions (see~\cite{kozlowski:gawelczyk:2002,allen:etal:2006}). The model described in Appendix~\ref{appendix:model:specification} naturally generalizes many of these models, and thus allows us to easily ask whether any of these simpler models are also adequate explanations of the empirical distribution.

\begin{figure*}[t]
\begin{center}
\includegraphics[scale=0.7]{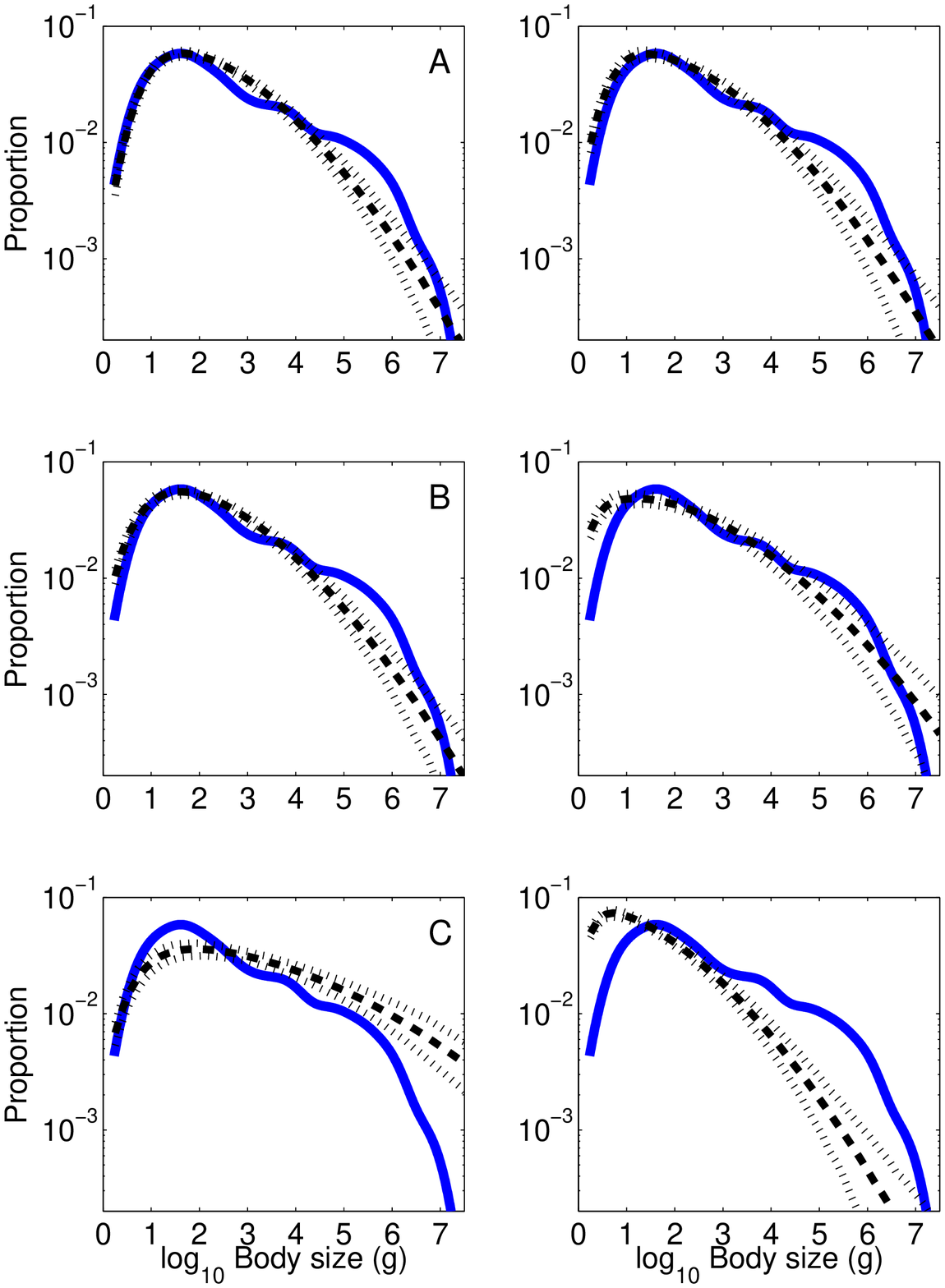} 
\end{center}
\caption{A comparison of the full model described in Appendix~\ref{appendix:model:specification} with several simpler models, all of which have a lower boundary $x_{\min}$. Results are presented in pairs, showing results for $F(\lambda)$ with power-law tails (left) and without (right). In all cases, model results show the central tendency of the model (over 1000 repetitions) with 95\% confidence intervals. (\textbf{A}) The full model as described in the text, with all parameters as set in Table~\ref{table:params}. (\textbf{B}) The same model as in \textbf{A}, but with no increase in $\langle\log\lambda\rangle$ for small-bodied species. (\textbf{C}) The same model as in \textbf{B}, but also with no size-dependent extinction risk and without Cope's rule for large-bodied species ($\langle\log\lambda\rangle=0$), i.e., a model of unbiased diffusion with a lower bound. Table~\ref{table:comparison} summarizes these results and gives the specific parameter settings used. }
\label{fig:comparison:1}
\end{figure*}

\begin{figure*}[t]
\begin{center}
\includegraphics[scale=0.7]{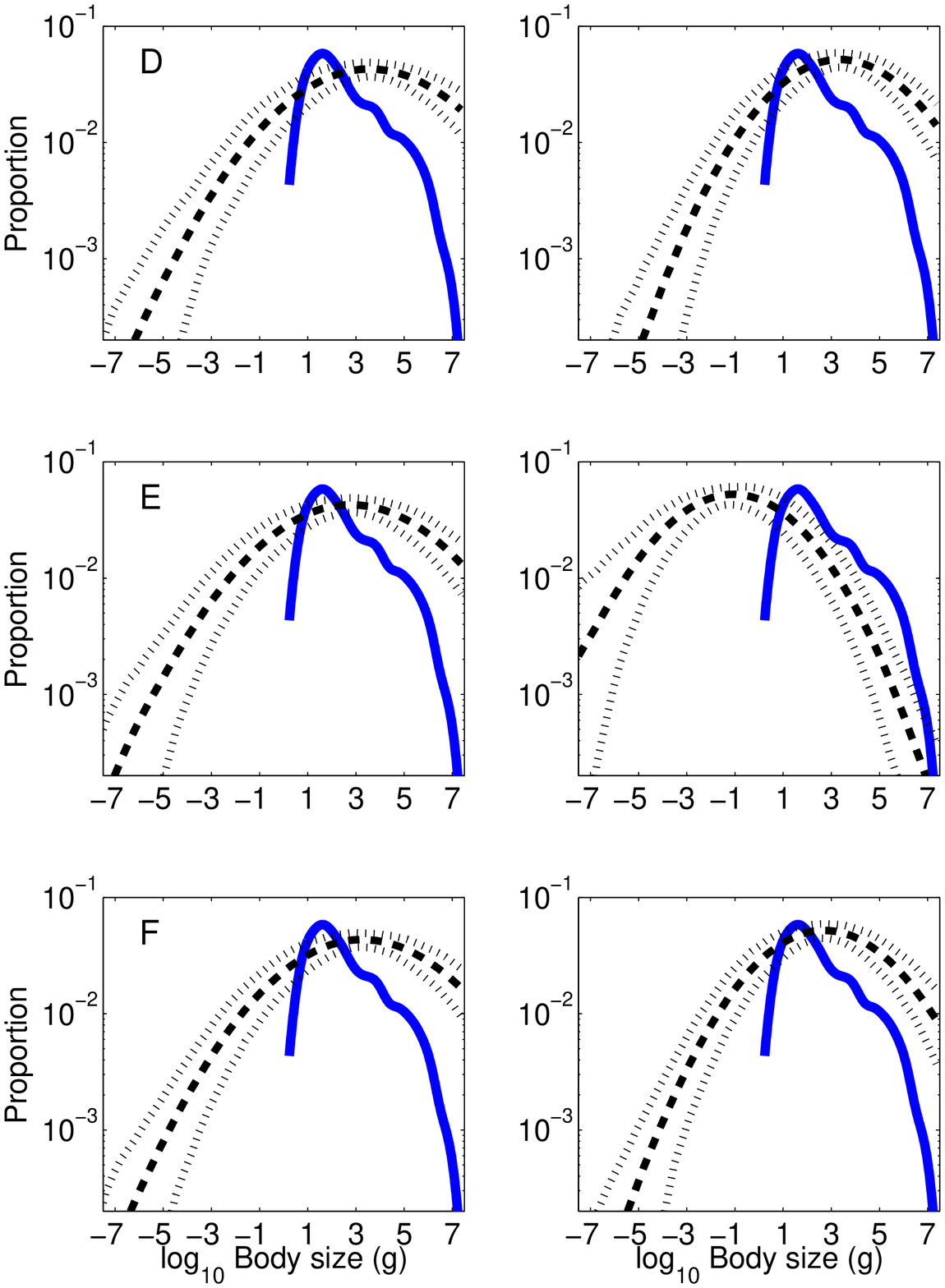} 
\end{center}
\caption{As in Fig.~\ref{fig:comparison:1}, results for simpler models are presented in pairs, showing $F(\lambda)$ with power-law tails (left) and without (right). All models shown here effectively have no lower boundary on size, i.e., we set $x_{\min}=10^{-8}\g$. (\textbf{D}) The model described in the text, but with no increase in $\langle\log\lambda\rangle$ for small-bodied species, i.e., a model with Cope's rule for all species and with size-dependent extinction risk. (\textbf{E}) The same model as in \textbf{D}, but with no size-dependent extinction risk, i.e., a model of Cope's rule alone. (\textbf{F}) The same model as in \textbf{D}, but without Cope's rule, i.e., a model with size-dependent extinction risk alone. Table~\ref{table:comparison} summarizes these results and gives the specific parameter settings used.}
\label{fig:comparison:2}
\end{figure*}

In particular, we consider (1) unbiased diffusion with a lower boundary, (2) Cope's rule with size-dependent extinction, (3) Cope's rule alone, and (4) size-dependent extinction alone. Additionally, we consider (5) a simplified version of the full model that omits the increased bias toward larger descendants for small-bodied species near $x_{\min}$, i.e., a model in which $\mu(x_{A})=\delta$ rather than the more complex form given in Eq.~\eqref{eq:muxa} (see Appendix~\ref{appendix:copesrule:model}). Simulation results for the full model and each of these five models are shown in Figs.~\ref{fig:comparison:1} and~\ref{fig:comparison:2}. The results of the experiments are summarized in Table~\ref{table:comparison}.

For each model, we repeated the simulation 1000 times to compute the simulated distributions' central tendencies (as in Fig.~\ref{fig:appendix:model}D). We also calculated the average distributional distance $\langle$wKS$\rangle$ from the empirical distribution, which we used to rank-order the models in terms of their accuracy. For models that included the mechanism for size-dependent extinction (i.e., when $\rho\not=0$), we re-estimated $\rho$ by minimizing $\langle$wKS$\rangle$ relative to the empirical distribution. In general, except as specified in Table~\ref{table:comparison}, the parameters of the included mechanisms were set according to our estimates from fossil data (Table~\ref{table:params}).

The results of this exercise indicate that the full model is the best explanation of the empirical distribution for terrestrial mammals ($\langle$wKS$\rangle=0.181$, Fig.~\ref{fig:comparison:1}A), reproducing the entire distribution quite accurately, with the exception of significant deviations near $1\kg$ and $300\kg$. We note that none of the alternative models could reproduce these deviations. Further, only the full model, which includes the increased bias for small-bodied species, accurately reproduces the left-tail of the empirical distribution. All other models, including the model that omits only this behavior but is otherwise identical to the full model (Fig.~\ref{fig:comparison:1}B), overestimate the number of species with size $x<40\g$. To be clear, the lower limit on body size itself causes the left-tail of the simulated distribution to decay somewhat like that of the empirical distribution, but only by including the increased bias for small-bodied species, inferred from fossil data (Appendix~\ref{appendix:copesrule:model}), do the tails coincide.

The second best model is the one that omits the small-size bias ($\langle$wKS$\rangle=0.244$, Fig.~\ref{fig:comparison:1}B). This model, however, fails to accurately reproduce the left-tail of the empirical distribution; the fit to the right-tail is largely unaffected. The third-best model is unbiased diffusion in the presence of a lower boundary but without a size-dependent extinction risk ($\langle$wKS$\rangle=2.97$, Fig.~\ref{fig:comparison:1}C). This model produces distributions with a heavy right-tail and a steep decline in density near $x_{\min}$, but dramatically misestimates the number of large-bodied species (too many for $F(\lambda)$ with power-law tail, and too few for $F(\lambda)$ with log-normal tails), and the number of species near the modal size $x\approx40\g$. This model also has the possibly undesirable feature of no steady-state. That is, the more time has passed, the heavier the distribution's right-tail, and the larger the largest extant mammal, becomes. This implies that the similarity of the simulated and empirical distributions, in this case, depends strongly on the mean species lifetime $\nu$ and the length of the \mbox{simulation $\tau$}.

The three models with no lower limit $x_{\min}$ failed to produce distributions remotely close to the empirical one, with $\langle$wKS$\rangle>10.6$ in all cases, and typically produced an over-abundance of extremely small species (e.g., $x<0.01\g$). It may be possible to improve these results by altering some model parameters far beyond the values estimated from fossil data, e.g., significantly increasing the strength of Cope's rule $\delta$ and the extinction risk at larger sizes $\rho$ to drive small-bodied species toward larger sizes. Alternatively, more complex mechanisms may also improve the results of these simple models, e.g., an extinction-risk curve that increases weakly above, and strongly below, $x\approx40\g$ would partly mimic the effect of a hard lower limit; using a more complex $F(\lambda)$ can certainly produce apparently complicated distributions (e.g.,~\cite{wang:2005}); etc.

Thus, all three processes -- a fundamental lower limit, the diffusion of species size, and an increasing risk of extinction with size -- are necessary to reproduce the empirical distribution of Recent terrestrial mammals, and models that omit either the lower limit $x_{\min}$ or extinction risks that increase with body size never produce realistic distributions, when using parameter estimates drawn from fossil data. Further, we found that an increased bias toward larger sizes for small-bodied species ($x\lesssim32\g$) is necessary to reproduce the particular shape of the empirical distribution's left-tail (small-bodied species); without this increased bias, the model consistently over-estimates the number of species near the lower-limit. Finally, we found that a systematic relationship between the strength of Cope's rule $\delta$ and the rate at which extinction risk increases $\rho$ is necessary to produce realistic body size distributions, such that an increase in the short-term benefit of increased size can be balanced by a comparable increase in the long-term risk of extinction from increased size.

\section{Simulation code}
\label{appendix:code}
This simulation code is written in the Matlab programming language. It requires no additional toolboxes to run, and should be compatible with all recent versions of the software.

\begin{verbatim}
% simulation parameters
xmin  = 1.8;   % lower bound
x0    = 40;    % founder body size
n     = 5000;  % num. species at equilbrium
beta  = 1/n;   % baseline extinction rate
rho   = 0.025; % rate of extinction increase
nu    = 1.6;   % mean species lifetime (My)
tau   = 60;    % total simulation time (My)
c(1)  = 0.33;  % log-lambda intercept
c(2)  = 1.30;  % log-size intercept
delta = 0.04;  % systematic bias (Cope's rule)
sigma = 0.63;  % variance
alpha = 0.30;  % power-law tail 

% data structure set up
tmax = round((tau/nu)*n);
xmax = 10^15;
x    = -Inf*ones(ceil(1.5*n),1);
x(1) = x0;
kdt  = 5000;
ns   = 1;
nk   = 0;
kd   = 1;
f_stop = 0; 

% begin main loop
while ~f_stop

     % begin cladogenesis step
     pair = [ceil(ns*rand(1)) ns+1];
     mass = x(pair(1),1);
     L1 = mass/xmin;  % lower bound
     L2 = xmax/mass;  % upper bound

     % model of Cope's rule
     if log10(mass)<c(2)
          % increased bias for small sizes
          mu = (-c(1)/c(2))* ...
               log10(mass)+c(1)+delta; 
     else
          % uniform bias for large sizes
          mu = delta;
     end; 
     
     % Monte Carlo draw of growth factors
     tt = [0 0];      
     while any(tt<1/L1 | tt>L2)
          % F(lambda) with power-law tails
          tt = exp(randn(2,1)*sigma+mu).* ...
          ((rand(2,1).* ...
               (1-1./L1)+1./L1).^alpha)./ ...
          ((rand(2,1).* ...
               (1-1./L2)+1./L2).^alpha);
     end;
     x(pair) = mass.*tt;
     kd = kd+2;
     ns = ns+1;
     % end cladogenesis step

     % begin extinction step
     % power-law model of extinction risk
     kl = rand(ns,1) < ...
          10.^(rho*log10(x(1:ns))+log10(beta));
     if sum(kl)>0
          x(1:sum(~kl)) = x(~kl);
          x(sum(~kl)+1:ns) = ...
               repmat([-Inf],sum(kl),1);
          ns = sum(~kl);
          nk = nk+sum(kl);
     end;
     % end extinction step

     % begin check stop-criteria
     if kd>=tmax, f_stop = 1; end;
     % end check stop-criteria
     
end;
% end main loop
\end{verbatim}

\end{appendix}


\begin{thebibliography}{10}

\bibitem{stanley:1973}
S.~M. Stanley, {\it Evolution\/} {\bf 27}, 1 (1973).

\bibitem{kozlowski:gawelczyk:2002}
J.~Koz{\l}owski, A.~T. Gawelczyk, {\it Functional Ecology\/} {\bf 16}, 419
  (2002).

\bibitem{allen:etal:2006}
C.~R. Allen, {\it et~al.\/}, {\it Ecology Letters\/} {\bf 9}, 630 (2006).

\bibitem{dial:marzluff:1988}
K.~P. Dial, J.~M. Marzluff, {\it Ecology\/} {\bf 69}, 1620 (1988).

\bibitem{brown:1995}
J.~H. Brown, {\it Macroecology\/} (University of Chicago Press, Chicago, 1995).

\bibitem{west:etal:2002}
G.~B. West, W.~H. Woodruff, J.~H. Brown, {\it Proceedings of the National
  Academy of Science, USA\/} {\bf 99}, 2473 (2002).

\bibitem{cardillo:etal:2005}
M.~Cardillo, {\it et~al.\/}, {\it Science\/} {\bf 309}, 1239 (2005).

\bibitem{fisher:owens:2004}
D.~O. Fisher, I.~P.~F. Owens, {\it Trends in Ecology and Evolution\/} {\bf 19},
  391 (2004).

\bibitem{stanley:1975}
S.~M. Stanley, {\it Proceedings of the National Academy of Science, USA\/} {\bf
  72}, 646 (1975).

\bibitem{sebens:1987}
K.~P. Sebens, {\it Annual Review of Ecology and Systematics\/} {\bf 18}, 371
  (1987).

\bibitem{brown:etal:1996}
J.~H. Brown, P.~A. Marquet, M.~L. Taper, {\it American Naturalist\/} {\bf 147},
  1092 (1996).

\bibitem{lomolino:1985}
M.~Lomolino, {\it American Naturalist\/} {\bf 125}, 310 (1985).

\bibitem{deperet:1909}
C.~Deperet, {\it The transformations of the animal world\/} (D. Appleton and
  Co., New York, 1909).

\bibitem{alroy:1998}
J.~Alroy, {\it Science\/} {\bf 280}, 731 (1998).

\bibitem{kozlowski:2002}
J.~Koz{\l}owski, {\it Functional Ecology\/} {\bf 16}, 540 (2002).

\bibitem{vanvalen:1973}
L.~{Van Valen}, {\it Evolution\/} {\bf 29}, 87 (1973).

\bibitem{mcshea:1994}
D.~W. Mc{S}hea, {\it Evolution\/} {\bf 48}, 1747 (1994).

\bibitem{smith:etal:2003}
F.~A. Smith, {\it et~al.\/}, {\it Ecology\/} {\bf 84}, 3403 (2003). {MOM}
  {V}ersion 3.6.1.

\bibitem{alroy:2008}
J.~Alroy, North {A}merican {F}ossil {M}ammal {S}ystematics {D}atabase (2008).
  Paleobiology Database Online Systematics Archive 3, {\tt
  http://paleodb.org/}.

\bibitem{mckinney:1990}
M.~L. Mc{K}inney, {\it Evolutionary Trends\/}, K.~J. Mc{N}amara, ed.
  (University of Arizona Press, 1990), pp. 75--118.

\bibitem{fortelius:2003}
M.~{Fortelius (coordinator)}, Neogene of the {O}ld {W}orld {D}atabase of
  {F}ossil {M}ammals ({NOW}) (2003). {U}niversity of {H}elsinki, {NOW} public
  release 030717, {\tt http://www.helsinki.fi/science/now/}.

\bibitem{pearson:1948}
O.~P. Pearson, {\it Science\/} {\bf 108}, 44 (1948).

\bibitem{smith:etal:2004}
F.~A. Smith, {\it et~al.\/}, {\it American Naturalist\/} {\bf 163}, 672 (2002).

\bibitem{valkenburgh:etal:2004}
B.~{Van Valkenburgh}, X.~Wang, J.~Damuth, {\it Science\/} {\bf 306}, 101
  (2004).

\bibitem{ludwig:1996}
D.~Ludwig, {\it Ecological Applications\/} {\bf 6}, 1067 (1996).

\bibitem{liow:etal:2008}
L.~H. Liow, {\it et~al.\/}, {\it Proceedings of the National Academy of
  Science, USA\/} {\bf 105}, 6097 (2008).

\bibitem{carrano:2006}
M.~T. Carrano, {\it Amniote Paleobiology\/}, M.~T. Carrano, T.~J. Gaudin, R.~W.
  Blob, J.~R. Wible, eds. (University of Chicago Press, 2006), pp. 225--268.

\bibitem{boas:2006}
M.~L. Boas, {\it Mathematical Methods in the Physical Sciences\/} (John Wiley
  \& Sons, Inc., Hoboken, NJ, 2006), third edn.

\bibitem{warton:etal:2006}
D.~Warton, I.~Wright, D.~Falster, M.~Westoby, {\it Biological Reviews\/} {\bf
  81}, 259 (2006).

\bibitem{erwin:2006}
D.~H. Erwin, {\it Annual Review of Earth and Planetary Sciences\/} {\bf 34},
  569 (2006).

\bibitem{press:etal:1992}
W.~H. Press, S.~A. Teukolsky, W.~T. Vetterling, B.~P. Flannery, {\it Numerical
  Recipes in {C}: {T}he Art of Scientific Computing\/} (Cambridge University
  Press, Cambridge, UK, 1992).

\bibitem{macfadden:1986}
B.~Mac{F}adden, {\it Paleobiology\/} {\bf 12}, 355 (1986).

\bibitem{jablonski:1997}
D.~Jablonski, {\it Nature\/} {\bf 385}, 250 (1997).

\bibitem{maurer:1998}
B.~Maurer, {\it Evolutionary Ecology\/} {\bf 12}, 925 (1998).

\bibitem{bokma:2002}
F.~Bokma, {\it Evolution\/} {\bf 56}, 2499 (2002).

\bibitem{novackgottshall:lanier:2008}
P.~M. Novack-Gottshall, M.~A. Lanier, {\it Proceedings of the National Academy
  of Science, USA\/} {\bf 105}, 5430 (2008).

\bibitem{reed:jorgensen:2004}
W.~J. Reed, M.~Jorgensen, {\it Communications in Statistics: Theory \&
  Methods\/} {\bf 33}, 1733 (2004).

\bibitem{clauset:etal:2007}
A.~Clauset, C.~R. Shalizi, M.~E.~J. Newman, Power-law distribution in empirical
  data (2007). E-print, {\tt arxiv:0706.1062}.

\bibitem{efron:tibshirani:1993}
B.~Efron, R.~J. Tibshirani, {\it An Introduction to the Bootstrap\/} (Chapman
  \& Hall, New York, NY, 1993).

\bibitem{vuong:1989}
Q.~H. Vuong, {\it Econometrica\/} {\bf 57}, 307 (1989).

\bibitem{wasserman:2006}
L.~Wasserman, {\it All of Nonparametric Statistics\/} (Springer, New York, NY,
  2006).

\bibitem{savage:etal:2004}
V.~M. Savage, {\it et~al.\/}, {\it Functional Ecology\/} {\bf 18}, 257 (2004).

\bibitem{wang:2005}
S.~C. Wang, {\it Paleobiology\/} {\bf 31}, 191 (2005).

\end{thebibliography}
\end{document}